\documentclass[a4paper,11pt]{article}
\pdfoutput=1 

\usepackage{jcappub} 
\usepackage[T1]{fontenc} 
\usepackage{color}
\usepackage{subfigure, wrapfig}

\graphicspath{{./Figures/final/}}

\title{\boldmath Sensitivity of Next-Generation Tritium Beta-Decay Experiments for keV-Scale Sterile Neutrinos }

\author[a,d]{S.\ Mertens}
\author[b,c]{T.\ Lasserre}
\author[d]{S.\ Groh}
\author[d]{G.\ Drexlin}
\author[d,f]{F.\ Gl\"{u}ck }
\author[d]{A.\ Huber}
\author[a]{A.\ W.\ P.\ Poon}
\author[d]{M.\ Steidl}
\author[e]{N.\ Steinbrink}
\author[e]{C.\ Weinheimer}

\affiliation[a]{Institute for Nuclear and Particle Astrophysics, Lawrence Berkeley National Laboratory, Berkeley, CA 94720, USA}
\affiliation[b]{Commissariat a l'\'energie atomique et aux \'energies alternatives, Centre de Saclay, IRFU, 91191 Gif-sur-Yvette, France}
\affiliation[c]{Astroparticules et Cosmologie APC, 10 rue Alice Domon et L\'eonie Duquet, 75205 Paris cedex 13, France}
\affiliation[d]{KCETA, Karlsruhe Institute of Technology, 76021 Karlsruhe, Germany}
\affiliation[e]{Institut f\"{u}r Kernphysik, Westf\"{a}lische Wilhelms-Universit\"{a}t M\"{u}nster, 48149 M\"{u}nster, Germany}
\affiliation[f]{Wigner Research Institute for Physics, P.O.B. 49, H-1525 Budapest, Hungary}

\emailAdd{smertens@lbl.gov}
\emailAdd{thierry.lasserre@cea.fr}
\emailAdd{stefan.groh@kit.edu}
\emailAdd{anton.huber@student.kit.edu}
\emailAdd{guido.drexlin@kit.edu}
\emailAdd{ferenc.glueck@kit.edu}
\emailAdd{awpoon@lbl.gov}
\emailAdd{markus.steidl@kit.edu}
\emailAdd{n.steinbrink@uni-muenster.de}
\emailAdd{weinheimer@uni-muenster.de}

\abstract{We investigate the sensitivity of tritium $\beta$-decay experiments for keV-scale sterile neutrinos. Relic sterile neutrinos in the keV mass range can contribute both to the cold and warm dark matter content of the universe. This work shows that a large-scale tritium beta-decay experiment, similar to the KATRIN experiment that is under construction, can reach a statistical sensitivity of the active-sterile neutrino mixing of $\sin^2\theta \sim 10^{-8}$. The effect of uncertainties in the known theoretical corrections to the tritium $\beta$-decay spectrum were investigated, and found not to affect the sensitivity significantly. It is demonstrated that controlling uncorrelated systematic effects will be one of the main challenges in such an experiment.}

\begin{document}
\maketitle

\section{Introduction}
During the past two decades, neutrino experiments have provided compelling evidence for neutrino mass through the discovery of neutrino flavor oscillations~\cite{NO1, NO2, NO3, pdg}.  The well-established standard neutrino oscillation framework comprises three light active neutrino mass eigenstates. However, there is a broad phenomenology addressing additional neutrino mass eigenstates~\cite{sn-whitepaper}. These new states would be predominantly sterile~\cite{snu-intro} (i.\ e. would not take part in Standard Model interactions) but could have a small admixture of active neutrinos~\footnote{Even though the additional heavy neutrino mass eigenstates are not exclusively sterile, we refer to them as (keV-scale) sterile neutrinos throughout this paper.}.

Very heavy sterile neutrinos are commonly postulated in see-saw mechanisms to explain the tiny masses of the light neutrinos~\cite{neutrino-masses}. Recently, there has been renewed interest in light sterile neutrinos due to puzzling experimental anomalies that could be explained by the existence of neutrinos with a mass in the eV range~\cite{sn-whitepaper}. In this article, we consider keV-scale sterile neutrinos which could account for a significant fraction of the dark matter in the universe~\cite{Dodelson1994,Shi99,Abaz01,Asaka05,Kusenko06,Merle2013,Drew13}. 

Recent results from the Planck satellite confirm that the universe is composed of 68.3\% dark energy, 26.8\% dark matter, and 4.9\% baryonic matter~\cite{Planck2013}. The nature of dark matter is one of the most intriguing questions of modern physics, since the Standard Model of elementary particle physics (SM) does not provide a suitable dark matter candidate. Such a candidate should be electrically neutral, at most weakly interacting, and stable with respect to the age of the universe. 

Relic active neutrinos, forming hot dark matter (HDM), are firmly ruled out as the dominant dark matter component. At the time of structure formation, light neutrinos had relativistic velocities and a large free streaming length, leading to a washing-out of small-scale structures, which is in disagreement with observations~\cite{Han2010, Mena2012}. Consequently, the most most favored candidate was thought to be a cold dark matter (CDM) particle, the so-called weakly interacting massive particle (WIMP). Its freeze-out in the early universe occurs at non-relativistic velocities, preventing the washing-out of small-scale structures. Furthermore, the existence of WIMPs is independently motivated by theories extending the SM, such as Supersymmetry~\cite{SusyDM}. WIMPs are actively sought in direct and indirect measurements, but no solid evidence for their existence yet has been reported~\cite{lux14}.

Relic sterile neutrinos, with a mass in the keV range, are a candidate for both warm and cold dark matter (WDM and CDM)~\cite{Dodelson1994,Shi99,Abaz01,Asaka05,Kusenko06,Merle2013, Drew13}. WDM and CDM scenarios fit the large-scale structure data equally well~\cite{Yang2013}. On the galactic scale WDM scenarios predict a smaller number of dwarf satellite galaxies and shallower galactic density profiles than CDM, resolving tensions between observations of galaxy-size objects and specific CDM model simulations~\cite{Klypin1999, Moore1999, Zavala2009, Tikhonov2009, deBlok2001, Vega2013, Lovell2012, San10, San12, San13, San14}.

Astrophysical observations constrain the sterile neutrino mass $m_{\mathrm{s}}$ and active-sterile mixing angle $\theta$. A robust and model-independent lower bound on the mass of spin-one-half dark matter particles is derived by considering the phase-space density evolution of dwarf spheroidal satellites in the Milky Way, leading to a mass limit of $m_{\mathrm{s}}>$1~keV~\cite{Boyarsky2009, Boyarsky2009b}. Another sensitive observable is the X-ray emission line at half of the neutrino mass, arising from the decay of a keV-scale sterile neutrino into an active neutrino and a photon, which can be searched for with appropriate X-ray Space Telescopes, such as XMM-Newton \cite{Watson2012} and Chandra~\cite{Boyarsky2008}. A combination of all observables lead to stringent but model-dependent bounds of 1~keV~$<m_{\mathrm{s}}<$~50~keV and 10$^{-13}<\sin^2(2\theta)<10^{-7}$~\cite{Boyarsky2013}. 

Interestingly, a recent analysis of the XMM-Newton telescope data might show an indirect evidence for the existence of relic neutrinos with a mass of $m_{\mathrm{s}}=7.1~\mathrm{keV}$ and an active-sterile neutrino mixing amplitude of $\sin^2\theta=7\cdot 10^{-11}$, based on indications of a very weak emission line from stacked galaxy cluster X-ray data~\cite{Bulbul}. A similar result has been reported for X-ray spectra of the M31 galaxy and the Perseus cluster~\cite{M31}. Such a neutrino, not excluded by any other data, could have been produced through a resonant Shi-Fuller mechanism~\cite{Abazajian2014}. 

In this work we address the prospects of searching for keV-scale sterile neutrinos with the next generation of $\beta$-decay experiments. A sterile neutrino would manifest itself as a ``kink''-like signature and a spectral distortion in the $\beta$-decay spectrum~\cite{Sch80, San13trit, Rod14, Rod14b}. In particular, tritium $\beta$ decay provides distinct advantages. First, tritium $\beta$-decay is of super-allowed type, and therefore a precise theoretical description of the spectral shape is possible. Second, the 12.3-year half-life of tritium is relatively short, allowing for high signal rates with low source densities, which in turn minimizes source-related systematic effects such as inelastic scattering. Finally, with an endpoint energy of $E_0=18.575$~keV, the phase space of tritium provides access to a search for heavy sterile neutrinos in a mass range of astrophysical interest. 

Currently, the Karlsruhe Tritium Neutrino Experiment (KATRIN) is the only large-scale tritium $\beta$-decay experiment planned to begin taking data in the near future \cite{KAT04, Drex13}. Although the primary mission of KATRIN is to investigate the narrow region ($\sim$30~eV) close to $E_0$, the unique properties of its tritium source characteristics allow, in principle, a  high-sensitivity search for sterile neutrinos by extending the measurements to the entire $\beta$-decay energy spectrum. In this paper, we qualitatively discuss the advantages and disadvantages of the KATRIN apparatus with respect to a sterile neutrino search and we gear our sensitivity studies to the expected tritium source strength of KATRIN as a benchmark point.

\section{Tritium Beta Decay and Sterile Neutrinos}
In this section we give a brief introduction to the generic shape of the tritium $\beta$-decay spectrum and describe the characteristic imprint of active-to-sterile neutrino mixing on the spectral shape. 

\subsection{Tritium Beta-Decay Spectrum}
We consider the super-allowed $\beta$ decay of molecular tritium
\begin{equation}
\mathrm{T}_2 \rightarrow  \mathrm{T}\mathrm{He}^{+} + e + \overline{\nu}_e.
\end{equation}
The $\beta$-decay spectrum of tritium is mainly governed by the phase space factor $p E^{\mathrm{tot}}p_{\nu}E_{\nu}^{\mathrm{tot}}$, where $p$, $p_{\nu}$, $E^{\mathrm{tot}}$ and $E^{\mathrm{tot}}_{\nu}$ are the electron and neutrino momentum and their total energy, respectively. The dominant correction to the simple phase space factor is the Coulomb interaction between the outgoing electron and the daughter nucleus, described by the Fermi function $F(E, Z=2)$. 

For a single neutrino mass eigenstate $m_{\nu}$ the decay rate is given by
\begin{equation}
\label{eq:betaspectrum1}
 \frac{d\Gamma}{dE} = C \cdot F(E, Z=2) \cdot p \cdot (E + m_e) \cdot (E_0 - E) \sqrt{(E_0 - E)^2 - m_{\nu}^2},
\end{equation}
where $E$ denotes the electron kinetic energy, $E_0$ is the endpoint energy for zero neutrino mass, $m_e$ and $m_{\nu}$ are the electron and neutrino mass, respectively. The speed of light $c$, and the Planck constant $\hbar$, are set to unity. The normalization constant $C$ is given by
\begin{equation}
 C= \frac{G_F^2}{2\pi^3}\cos^2\Theta_C | M|^2,
\end{equation}
with $G_F$ denoting the Fermi constant, $\Theta_C$ the Cabbibo angle, and $M$ the energy-independent nuclear transition matrix element. A fully relativistic computation of the tritium $\beta$-decay spectrum in the elementary particle treatment of the weak interaction can be found in~\cite{Sim08,Mas07}.

In tritium $\beta$ decay an electron neutrino flavor eigenstate is created, which is a superposition of mass eigenstates. The spectrum is therefore a superposition of the spectra corresponding to each mass eigenstate $m(\nu_i)$, weighted by its mixing amplitude $|U_{ei}|$ to the electron flavor,
\begin{equation}
\label{eq:betaspectrum2}
 \frac{d\Gamma}{dE} = C \cdot F(E,Z=2) \cdot p \cdot (E + m_e) \cdot (E_0 -E) \sum_i | U_{ei} |^2 \sqrt{(E_0-E)^2 - m(\nu_i)^2}.
\end{equation}
Since the mass splittings between the three light mass eigenstates are so small, no current $\beta$-decay experiment can resolve them. Instead, a single effective light neutrino mass $m_{\mathrm{light}}^2 = \sum_{i=1}^{3} | U_{ei} |^2 m(\nu_i)^2$ is assumed. 

\subsection{Imprint of Active-Sterile Neutrino Mixing on the Beta-Decay Spectrum}
\begin{figure}
  \centering
  \subfigure[]{\includegraphics[width = 0.49\textwidth]{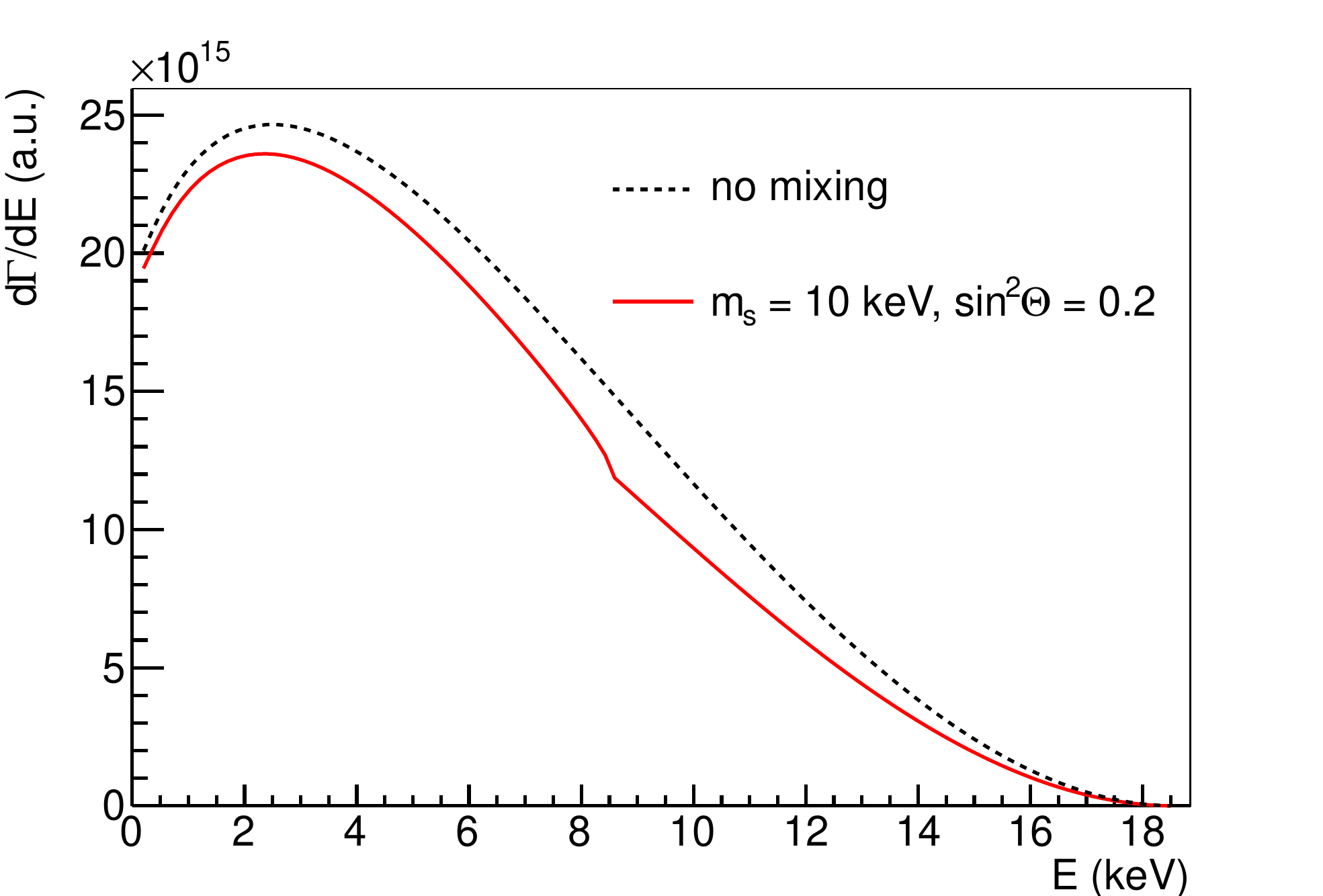}}
  \subfigure[]{\includegraphics[width = 0.49\textwidth]{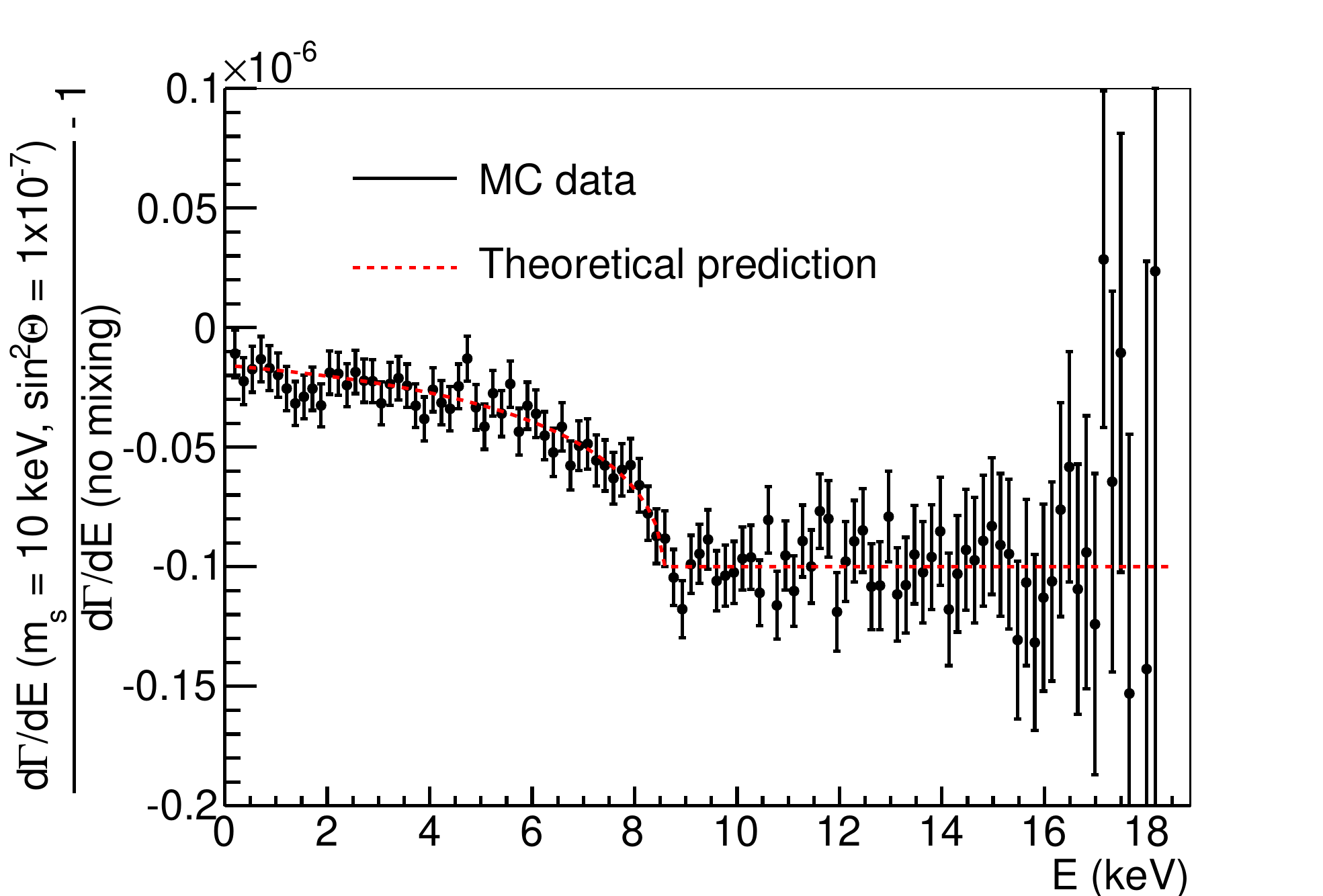}}
  \caption{a: A tritium $\beta$-decay spectrum with no mixing (dashed black line) compared to a spectrum with a keV sterile neutrino mass of 10~keV and a mixing angle of $\sin^{2}\theta=0.2$ (solid red line). One can clearly see a kink-like signature of the keV-scale sterile neutrino at the electron energy $E = E_0 - m_{\mathrm{s}}$ and its influence on the spectral shape below the kink energy. b: Ratio of a tritium $\beta$-decay spectrum without mixing and a spectrum with a 10~keV neutrino mass and a mixing amplitude of $\sin^{2}\theta=10^{-7}$. The error bars correspond to a total statistics of $\sim 10^{18}$ electrons, as would be achieved with the KATRIN tritium source after 3 years of measurement time.}
 \label{fig:Kink}
\end{figure}

If the electron neutrino contains an admixture of a neutrino mass eigenstate with a mass $m_{\mathrm{s}}$ in the keV range, the different mass eigenstates will no longer form one effective neutrino mass term. In this case, due to the large mass splitting, the superposition of the $\beta$-decay spectra corresponding to the light effective mass term $m_{\mathrm{light}}$ and the heavy mass eigenstate $m_{\mathrm{s}}$, can be detectable. In the following analysis we assume a single heavy mass eigenstate $m_{\mathrm{s}}$ and an effective light neutrino mass eigenstate $m_{\mathrm{light}}$. In this scenario the differential spectrum can be written as
\begin{equation}
\label{eq:tritiumspec}
 \frac{d\Gamma}{dE} = \cos^2\theta \frac{d\Gamma}{dE}(m_{\mathrm{light}}) + \sin^2\theta \frac{d\Gamma}{dE}(m_{\mathrm{s}}),
\end{equation} 
where $\theta$ describes the active-sterile neutrino mixing, and predominantly determines the size of the effect on the spectral shape. The light neutrino mass $m_{\mathrm{light}}$ is set to zero in our analysis, as $m_{\mathrm{s}}\gg m_{\mathrm{light}}$ is assumed. Figure~\ref{fig:Kink}a shows for an (unrealistically) large mixing angle of $\theta \approx 26^{\circ}$, how an admixture of a heavy mass eigenstate of $m_{\mathrm{s}} = 10$~keV manifests itself in the shape of the tritium $\beta$-decay spectrum. 

In the energy region $E < E_0 - m_{\mathrm{s}}$, the tritium $\beta$ decay into a keV-scale sterile neutrino is energetically allowed, leading to a kink in the spectrum at $E_{\mathrm{kink}} = E_0 - m_{\mathrm{s}}$. Due to its large mass, a keV-scale sterile neutrino is still non-relativistic at energies of a few keV, resulting in a characteristic spectral modification in the electron energy range up to $E < E_{\mathrm{kink}}$.

In the case of very small mixing, it is helpful to visualize the effect of a keV-scale sterile neutrino as the ratio of the spectrum with and without mixing, as shown in figure~\ref{fig:Kink}b. Taking into account the statistical error bars expected for a total statistics of $\sim 10^{18}$ electrons (as expected from the KATRIN tritium source, see section \ref{sec:KATRIN}), a small mixing amplitude of $\sin^2\theta \approx 10^{-7}$ still leads to both a clearly visible kink signature and an extended spectral distortion for $E < E_{\mathrm{kink}}$. 

\section{A KATRIN-like Experiment as an Apparatus for keV-Scale Neutrino Search}
\label{sec:KATRIN}
The KATRIN experiment is a next-generation, large-scale, single $\beta$-decay experiment~\cite{KAT04, Drex13}. It is currently under construction at the Tritium Laboratory at the Karlsruhe Institute of Technology (KIT) and will prospectively start taking data in 2016. The key goal of the experiment is to probe the effective light neutrino mass $m_{\mathrm{light}}$ with a sensitivity of 200~meV at 90\% confidence level (CL) by analyzing the shape of the tritium $\beta$-spectrum in a narrow region below the $\beta$-decay endpoint energy, where the impact of the light neutrino mass is maximal. 

\begin{figure}[]
\begin{center}
\includegraphics[width = \textwidth]{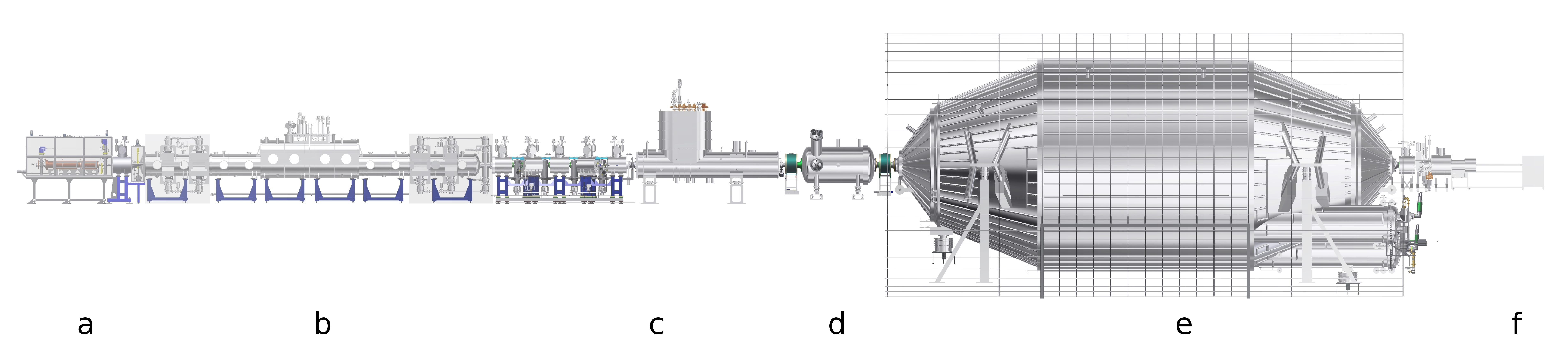}
\caption{Main components of the KATRIN experimental setup. a: rear section, b: windowless gaseous tritium source, c: differential and cryogenic pumping section, d: prespectrometer, e: main spectrometer, f: focal plane detector.}
\label{fig:KATRIN}
\end{center}
\end{figure}

Considering that only $10^{-13}$ of the $\beta$-decay electrons are created within an energy in the last 1~eV of the tritium $\beta$-decay spectrum, an extremely high decay rate is needed to reach the desired light neutrino mass sensitivity. KATRIN makes use of a gaseous molecular tritium source of very high activity ($\lambda_d \approx 1\cdot10^{11}$ decays per second) and stability (at the level of < $10^{-3}$ per hour). These unique source properties may allow KATRIN to extend its physics reach after having achieved its primary goal of measuring the light neutrino mass in the sub-eV range, to look for contributions of possible heavy neutrinos in the eV to multi-keV range. In the following we discuss the properties of the KATRIN main components (see figure~\ref{fig:KATRIN}) when searching for keV-scale sterile neutrinos in the entire tritium $\beta$-decay spectrum.

\subsection{Advantages and Limitations of KATRIN with Respect to a keV-Scale Neutrino Search}
The key component of KATRIN in view of the keV-scale sterile neutrino search is the high luminosity of its molecular tritium source. Its decay rate of $\lambda_d \approx 1\cdot10^{11}$ decays per second corresponds to a counting rate of $\lambda_r \approx 1.5\cdot10^{10}$~cps (taking into account acceptance angle and the transmitted cross section of the source) and a total statistics of $N_{\text{decays}} \approx 1.4 \cdot 10^{18}$ electrons after 3 years of measurement time. Beyond that, the KATRIN source features high isotopic purity (95\% T$_2$), which is constantly monitored by Laser Raman spectroscopy~\cite{Stu10, Sch11, Fis11}. With temperature variations much smaller than 30~mK and a precise monitoring of the tritium column density~\cite{Bab12, Gro11}, the decay rate is expected to be extremely stable. 

On the other hand, the gaseous tritium source entails a number of systematic effects. Energy losses of the $\beta$ electrons due to inelastic scattering in the source are unavoidable~\cite{Ass00}. Furthermore, the source section poses magnetic traps for electrons emitted under a large angle with respect to the beam axis. These initially trapped electrons can escape eventually by scattering into the beam. At the cost of a reduced statistics, systematic effects due to scattering may partly be reduced by lowering the source strength.

Another systematic effect related to the KATRIN setup is due to backscattering of $\beta$ electrons from the gold surface of the rear wall of the KATRIN setup~\cite{Bab12}. Placing the rear wall in a low magnetic field would be a possibility to reduce the backscattering and the returning of backscattered electrons into the source. For the keV-scale sterile neutrino search an in depth study of these energy-dependent source and rear wall effects based on the final design will be necessary. 

KATRIN provides a sophisticated tritium retention system, reducing the tritium flow by 14 orders of magnitude from the windowless gaseous tritium source (WGTS) to the spectrometer. This system is based on differential~\cite{Luk11} and cryogenic~\cite{Gil10, Luo08} pumping obviating any tritium blocking membrane. This fact is extremely advantageous for the search for keV-scale sterile neutrinos, as it avoids systematic effects due to energy losses of the electrons in such membrane. A scenario in which a detector would be installed at the rear section of KATRIN is disfavored since the tritium flow is not reduced as efficiently at this location.

The KATRIN main spectrometer works as a MAC-E filter, where a magnetic adiabatic collimation is combined with electrostatic filtering~\cite{KAT04}, providing an unsurpassed energy resolution of <~1~eV. The basic working principle is to apply a certain retarding potential to the spectrometer, which allows only electrons with a higher kinetic energy than this potential to pass and reach a counting detector. Hence, by recording the count rate for different retarding potentials the \textit{integral} shape of the energy spectrum is measured.

In the case of a keV-scale sterile neutrino search the entire tritium $\beta$-decay spectrum is of interest and therefore the main spectrometer would operate at very small retarding energies to allow the electrons of the interesting part of the spectrum to reach the detector. To guarantee an adiabatic transport of electrons with high surplus energy through the spectrometer~\cite{Pra12}, the magnetic field at the center of the spectrometer has to be increased by a factor of $\sim$10 as compared to normal KATRIN measuring mode. This can be achieved by making use of the large air coil system installed around the main spectrometer~\cite{Glu12}.

One of the biggest challenges in using the present KATRIN experimental setup as an apparatus to search for keV-scale sterile neutrinos arises from the high count rates of $\lambda_r = 1.5\cdot10^{10}$~cps when aiming to measure the entire $\beta$-decay spectrum. The 148 pixel silicon detector~\cite{Dun08}, as well as the electronics and DAQ systems, presently in use for normal KATRIN operation, are not designed to deal with these electron rates. 

Consequently, a new detector and read out system with the capability of handling extremely high count rates have to be designed. Depending on the measurement mode, a possible realization could be a large multi-pixel detector (assuming a count rate of 100~kHz per pixel is feasible, the detector would be composed of up to $10^{5}$ detector pixels). Promising technologies for high rate multi pixel detectors include silicon drift detector~\cite{Gatti84,Lech96}, MAPS (monolithic active pixel sensors)~\cite{Tur01} or DEPFET (depleted p-channel field effect transistor) systems~\cite{Nee02,Fis07}. The detector design has to be optimized in a way that allows for minimizing and quantifying numerous systematic uncertainties, including electron backscattering, energy loss in the dead layer, pileup, dead time, and energy nonlinearities. This paper is not intended to outline a detailed technical realization of a new detector system. In the following, we assume a hypothetical detector that is able to cope with the high count rates.

\subsection{Possible Measurement Schemes}
\label{subsec:upgrates}
\begin{figure}
  \centering
  \begin{minipage}{0.65\textwidth}
    \includegraphics[width = \textwidth]{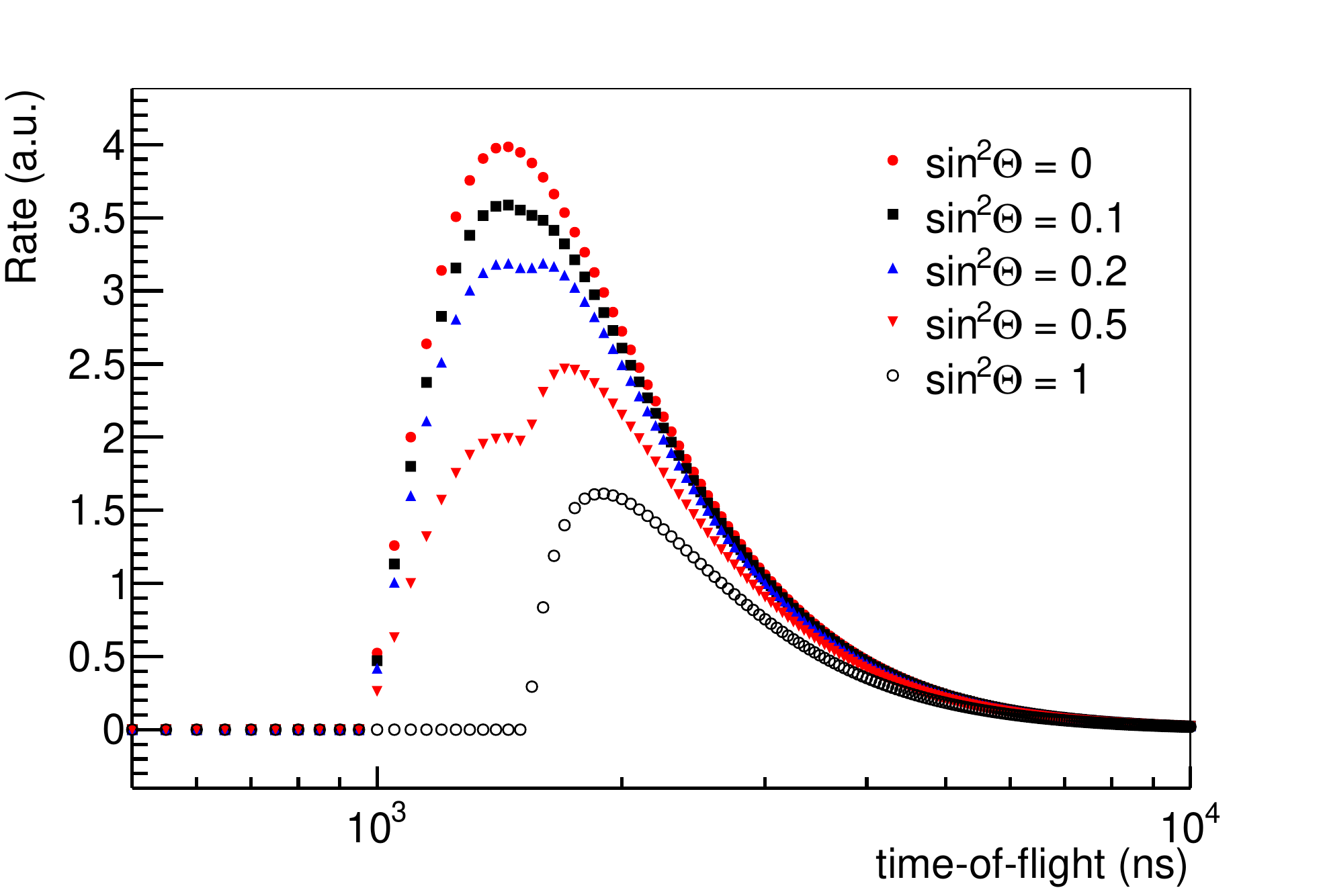}
  \end{minipage}
  \hfill
  \begin{minipage}{0.34\textwidth}
	\caption{ Time-of-flight tritium spectra for a keV-scale sterile neutrino of $m_{\mathrm{s}} = 1$~keV and different mixing angles. The mixing angles are unrealistically large to enhance the signature for the reader. The spectra have been computed for a fixed retarding potential of 17~keV. As in the tritium $\beta$-decay energy spectrum, the signature of a 1~keV neutrino occurs as a sudden change in the spectral shape of the time-of-flight spectrum.}
	\label{fig:tof}
  \end{minipage}
\end{figure}

Three general modes of operating a KATRIN-like experiment for the keV-scale sterile neutrino search can be considered: 
\begin{itemize}
\item Method A: An integral measurement, as in the standard KATRIN operation, could be performed, by measuring the count rate for different retarding potentials of the main spectrometer. In this case, the detector system only needs to meet the requirements with respect to high rate. The spectrometer would provide an ultra-sharp energy cut-off.
\item Method B: A differential measurement could be realized by using a detector system which itself provides a sufficient energy resolution. In this case, the spectrometer would be set to a fixed low retarding potential at all times, allowing the entire part of the spectrum of interest to reach the detector. The requirements on the detector system are technologically more challenging in this case. (If required and feasible, the detector system could also placed directly behind the last tritium retention system, thus bypassing the tandem spectrometer section and systematic effects associated to possible non-adiabatic transport of electrons through the MAC-E filters.)
\item Method C: Finally, an interesting method for the keV-scale sterile neutrino search could be a hybrid method based on a measurement of the time-of-flight of the electrons, as described in~\cite{Ste13}. Using the pre-spectrometer to define a sharp interval for the start time by the gated-filter method~\cite{Ste13} and the detector to define the stop time, it is possible to determine the time-of-flight of the 'slow' electrons whose kinetic energy is slightly higher than the retarding potential of the main spectrometer. The resolution of the time-of-flight spectrum is defined by the KATRIN spectrometer and not by the detector. The time-of-flight can only be measured in a rather small energy interval above the retarding potential. If the signature of the keV-scale sterile neutrino is in this energy region, it is visible in the time-of-flight spectrum, see figure~\ref{fig:tof}. By setting the spectrometer to different retarding potentials, the full $\beta$-decay spectrum can be scanned.
\end{itemize}

\section{Expected Statistical Sensitivity of a Tritium Beta Decay Experiment}
\label{ssc:results}
In this section we consider the statistical sensitivity of a tritium beta decay experiment. We calculate the statistical sensitivity for different tritium source strength and compare the option of a differential and integral measurement mode. 

\subsection{Model of Measured Tritium Spectrum}
\label{subsec:statsen}
The $\beta$ electron detection rate in an energy interval $i$ between $E_1$ and $E_2$ is given by:
\begin{equation}
 D_i = C_T\int_{E_1}^{E_2}\frac{d\Gamma}{dE}dE,
\end{equation}
where $\frac{d\Gamma}{dE}$ is the differential tritium spectrum, as defined in equation~\ref{eq:tritiumspec}, and $C_T$ is a normalization constant. As a realistic benchmark point we use experimental parameters of KATRIN. In this case, $C_T$ incorporates 
the maximal acceptance angle $\Theta_{\mathrm{max}}=51^{\circ}$, and the number of tritium molecules in the tritium source. The latter is determined by the product of the isotopic purity $\epsilon_{\mathrm{T}} = 95\%$, the flux tube cross section $\sigma_{\mathrm{WGTS}}= 53$~cm$^2$, and the column density $\cal{N}$ = $5\cdot10^{17}~\mathrm{molecules}/\mathrm{cm}^{2}$. Incorporating these factors leads to an effective number of tritium atoms of
\begin{equation}
\label{eq:Neff}
 \mathrm{N}_{\mathrm{eff}}=2{\cal{N}}\sigma_{\mathrm{WGTS}} \cdot \epsilon_{\mathrm{T}} \cdot 0.5\left(1-\cos\left(\Theta_{\mathrm{max}}\right)\right)= 9.2 \cdot 10^{18}
\end{equation}

In case of a differential measurement by an energy-resolving detector, the total number of counts in the interval $i$ is given by the product of the detection rate $D_i$ and the total measurement time t. For this analysis we assume t = 3 years. For the statistical analysis we consider a binned $\beta$-decay spectrum subdivided into 100 bins from 0.1 -- 18.574~keV, corresponding to a bin width of 186.6~eV. In our analysis we set the endpoint to $E_0=18.575$~keV. In view of the high signal count rates, the expected background level of less than 1~cps is considered as negligible.

The integral spectrum is obtained by measuring the number of counts with the spectrometer operating at different retarding potentials $U_{\mathrm{ret}}$. The number of counts at a specific retarding potential is calculated by integrating the differential spectrum from $qU_{\mathrm{ret}}$ to the endpoint. In this analysis we assume the transmission function to be a step function, thereby neglecting the finite spectrometer energy resolution (which is in the eV range~\cite{KAT04,Drex13}). We choose 100 equally distributed retarding potentials, a total measurement time of 3 years and an equal measurement time at each potential setting.  

\subsection{Statistical Sensitivity}
The 90\%~exclusion contours are obtained by scanning the ($m_{\mathrm{s}}, \sin^2\theta$) parameter space, computing the $\chi^2$ value at each point of the grid and drawing the curve such that $\Delta \chi^2$=$\chi^2$-$\chi^2_{\mathrm{min}}$ < 4.60. The $\chi^2$ function is given by 
\begin{equation}
\label{chi2}
\chi^2 = \sum_i \frac{\left( N_{\mathrm{theo}}^{i}-N_{\mathrm{exp}}^{i}\right)^2}{\sigma^2_{i}},
\end{equation}
where $N_{\mathrm{theo}}^{i}$ are the simulated data in the no-mixing case, $N_{\mathrm{exp}}^{i}$ denote the expectations for a given mixing scenario involving a keV-scale sterile neutrino, and $\sigma_{i}$ are the statistical uncertainty in each energy bin~$E_i$.

\begin{figure}
  \centering
  \subfigure[]{\includegraphics[width = 0.49\textwidth]{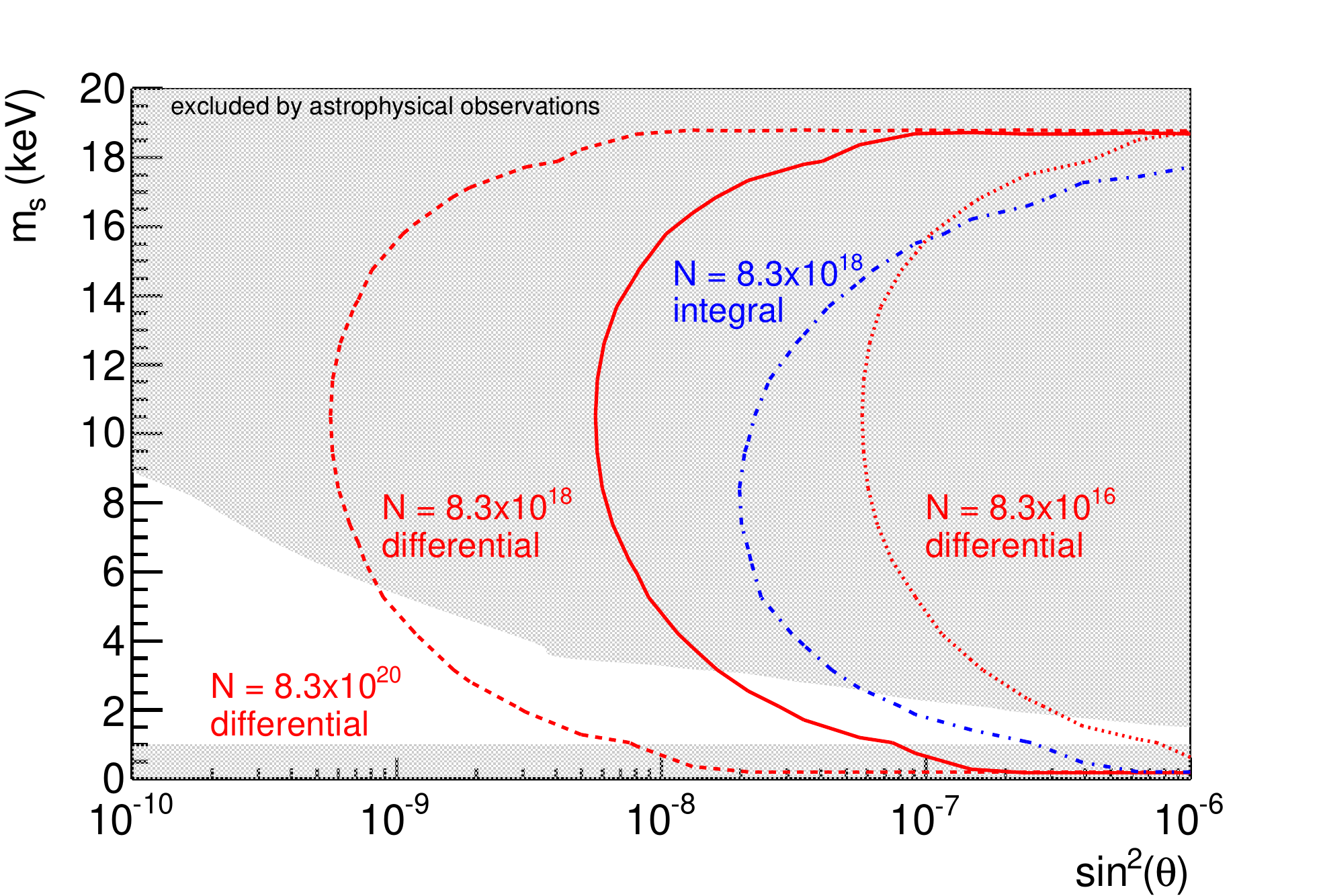}}
  \subfigure[]{\includegraphics[width = 0.49\textwidth]{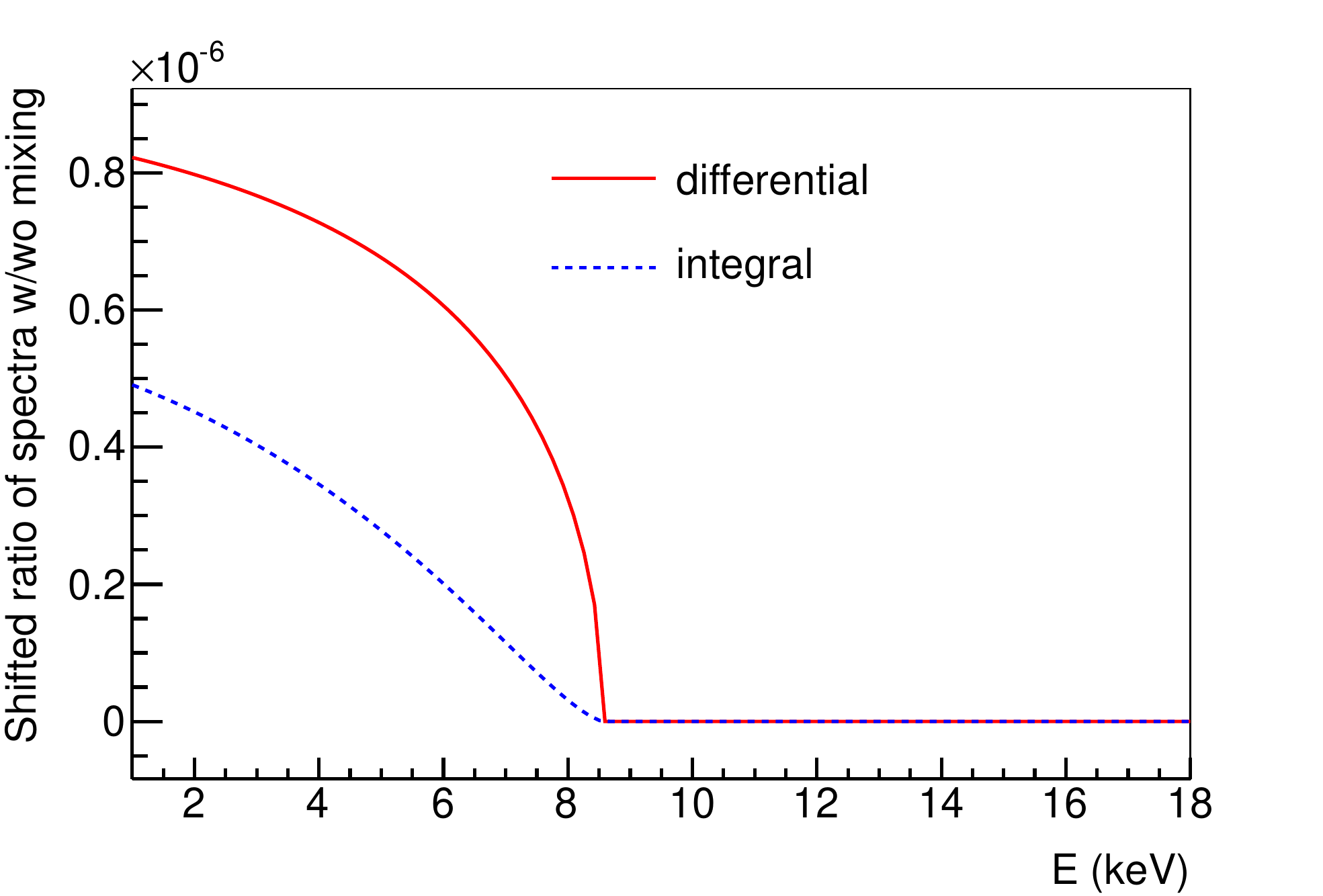}}
  \caption{a) 90\% statistical exclusion limit of a differential 3-years measurement with different effective numbers of tritium atoms in the source. With a generic detection efficiency of 90\%, the solid red line corresponds to the expected tritium source strength in KATRIN ($\mathrm{N} = 0.9 \cdot \mathrm{N}_{\mathrm{eff}}$, see equation~\ref{eq:Neff}). The blue dotted-dashed line displays the 90\% exclusion limit of an integral measurement of 3 years with the full KATRIN source strength. The gray shaded areas represent qualitatively the parameter space excluded by astrophysical observations~\cite{Drew13}. b) Ratio of spectra with $m_{\mathrm{s}}=10$~keV and $\sin^2\theta=10^{-6}$ and no mixing in the case of a differential (red solid line) and an integral (dashed blue line) measurement. In this representation both curves have been shifted to zero above the kink, to allow for an easy comparison.}
 \label{fig:DiffInt}
\end{figure}

In figure~\ref{fig:DiffInt}a we compare the statistical sensitivity for different tritium source strengths. A differential measurement with $\mathrm{N}_{\mathrm{eff}}= 9.2 \cdot 10^{18}$ (corresponding to the full KATRIN source strength) and a generic detection efficiency of 90\% reaches a statistical sensitivity  of $\sin^2\theta< 10^{-7}$ for a mass range of $m_{\mathrm{s}} \approx 1 - 18$~keV at 90\% CL, probing the cosmologically allowed region. To mitigate the requirements on the detector system and to partly reduce source-related systematic effects the source strength may need to be reduced. For a source luminosity reduced by a factor of 100 the statistical sensitivity at 90\% CL of a differential measurement still exceeds $\sin^2\theta = 10^{-7}$ for masses of $m_{\mathrm{s}} \approx 5 - 15$~keV. Considering the option of a new experimental approach, that would allow to increase the source strength by a factor of 100 a statistical sensitivity of $\sin^2\theta < 10^{-9}$ would be reached for a mass range of $m_{\mathrm{s}} \approx 5 - 16$~keV.  

For an integral measurement with $\mathrm{N}_{\mathrm{eff}}= 9.2 \cdot 10^{18}$ and a generic detection efficiency of 90\%, a statistical sensitivity of better than $\sin^2\theta = 10^{-7}$ can be achieved for masses of $m_{\mathrm{s}} \approx 2 - 15$~keV at 90\% CL. The sensitivity with an integral measurement is reduced for the following reason: As shown in figure~\ref{fig:DiffInt}b, the admixture of a keV neutrino mass eigenstate to the electron neutrino leads to a sudden increase in event rate at the energy $E = E_0-m_{\mathrm{s}}$. In an integral measurement the effect is reduced since the count rate at each measurement point includes all electrons with energies from the filter potential up to the endpoint. Therefore, the relative effect of a kink is the smallest far away from the endpoint, and consequently, a differential measurement is superior in the higher mass region, corresponding to a kink at small electron energies.  

Although, from a statistical point of view the differential measurement mode is preferable, the final sensitivity will depend on the details of the technical implementation of either method and the systematic effects associated with it. For instance, the differential measurement will be sensitive to any detector related effect, whereas an integral measurement would be prone to any instabilities of the retarding potential. A combination of both methods, accordingly, will be extremely valuable to probe a broad class of spectrometer and detector related systematic effects. The TOF method would especially be suited to cross-check a possible positive signal, as it allows to measure a specific small section of the spectrum with ultra-high energy resolution~\cite{Ste13}. To further evaluate the different measurement modes, one can artificially introduce a positive signal, as proposed by~\cite{Mor93}, using radioactive sources or an electron-gun and thereby test the capability of the methods to detect the characteristic signature.

\section{Impact of Theoretical Uncertainties}
In this section the impact of theoretical uncertainties in the calculation of the tritium $\beta$-decay spectrum is discussed. After reviewing the current state of the theoretical description of the tritium $\beta$-decay spectrum, we estimate the effect of theoretical uncertainties with the spectral fit approach.  By parameterizing the uncertainties of the spectral shape and letting these parameters vary in the fit, they may imitate the signature of a keV-scale sterile neutrino, and thereby alter the sensitivity.

\subsection{Theoretical Corrections to the Tritium Spectrum}
\label{sec:theoerror}
Theoretical corrections to the tritium $\beta$-decay spectrum arise at the particle, nuclear, atomic and molecular levels. In this paper we include the theoretical corrections to the tritium $\beta$-decay spectrum that are available in the literature. We discuss the missing information that will eventually be required for performing the spectral fit approach with a statistics of $10^{18}$ electrons. A summary of the status of the corrections is given in table~\ref{tab:corrections}. Figure~\ref{fig:allcorr} summarizes the considered correction terms. Their explicit expression is given in Appendix~\ref{appendix}. 

\begin{figure}[!h]
  \centering
  \includegraphics[width =.95\textwidth]{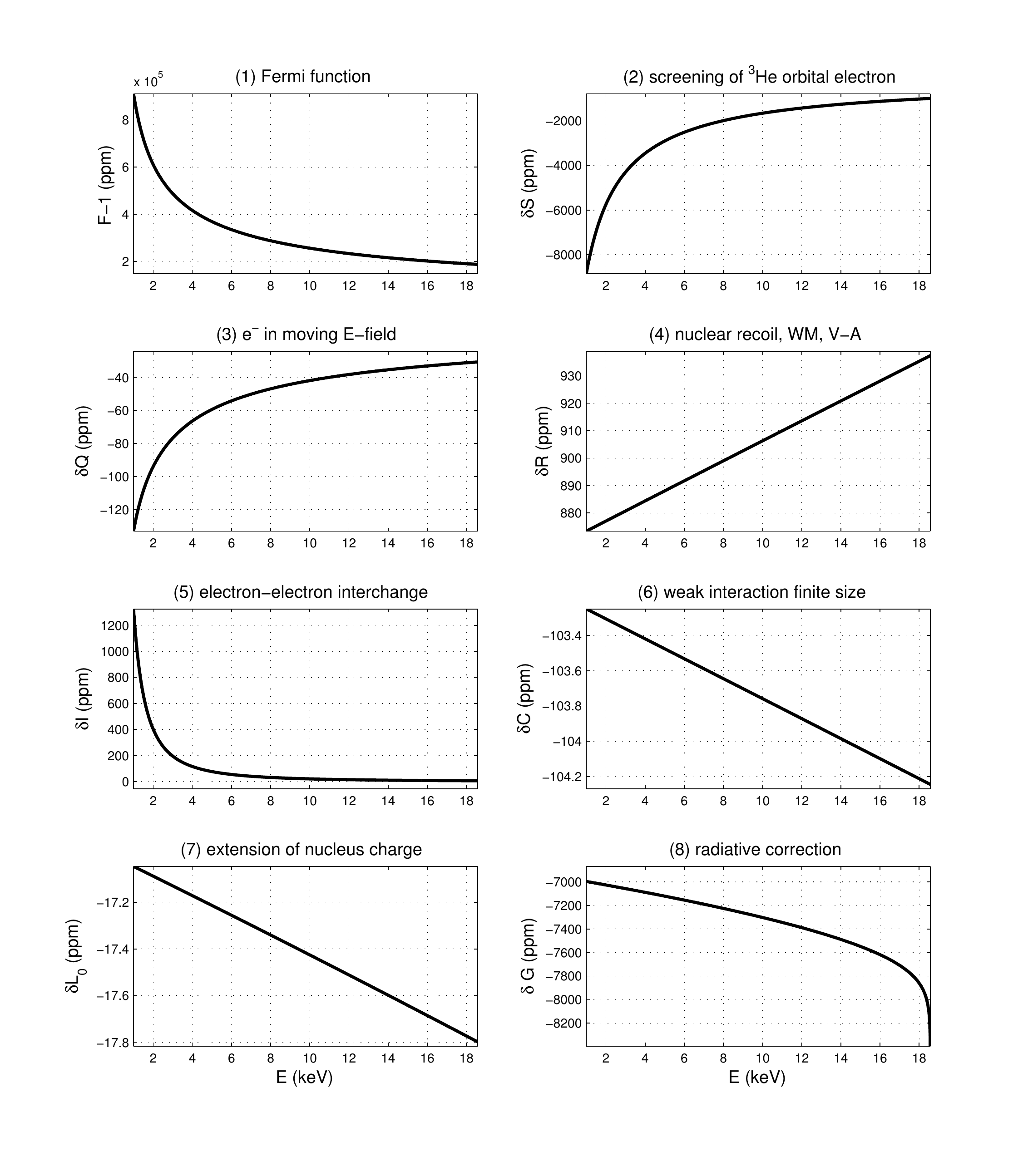}
  \hfill
  \caption{Relative difference $\delta \Psi_i=\frac{(d\Gamma/dE)^{\mathrm{corr}}}{(d\Gamma/dE)^{\mathrm{uncorr}}}-1$ ($\Psi_i$ = F, S, Q, R, I, C, L$_{0}$, G) of the tritium $\beta$-decay spectrum~\cite{Wilkinson1991} with $(d\Gamma/dE)^{\mathrm{corr}}$ and without $(d\Gamma/dE)^{\mathrm{uncorr}}$ theoretical corrections, expressed in parts per million (ppm) and displayed as a function of energy in the range $[E = 1~\mathrm{keV}; E = 18~\mathrm{keV}]$. In addition to the conventional relativistic Fermi function (1) the following effects are evaluated: (2) screening by the orbital electron \textbf{S};  (3) recoil of the He ion in its generation of the Coulomb field \textbf{Q}; (4) effect of the finite nuclear mass in the energy dependence of the phase space, V-A interference, and weak magnetism \textbf{R}; (5) interchange between the $\beta$-particle and the orbital electron \textbf{I}; (6) effect of the finite nuclear size on the solution of the Dirac equation for the electron \textbf{C}; (7) convolution of 
the lepton and nucleonic wave functions through the nuclear volume \textbf{L$_{0}$}; (8) first order radiative correction \textbf{G}.
}
\label{fig:allcorr}
\end{figure}

\begin{figure}
  \centering
  \subfigure[]{\includegraphics[width = 0.49\textwidth]{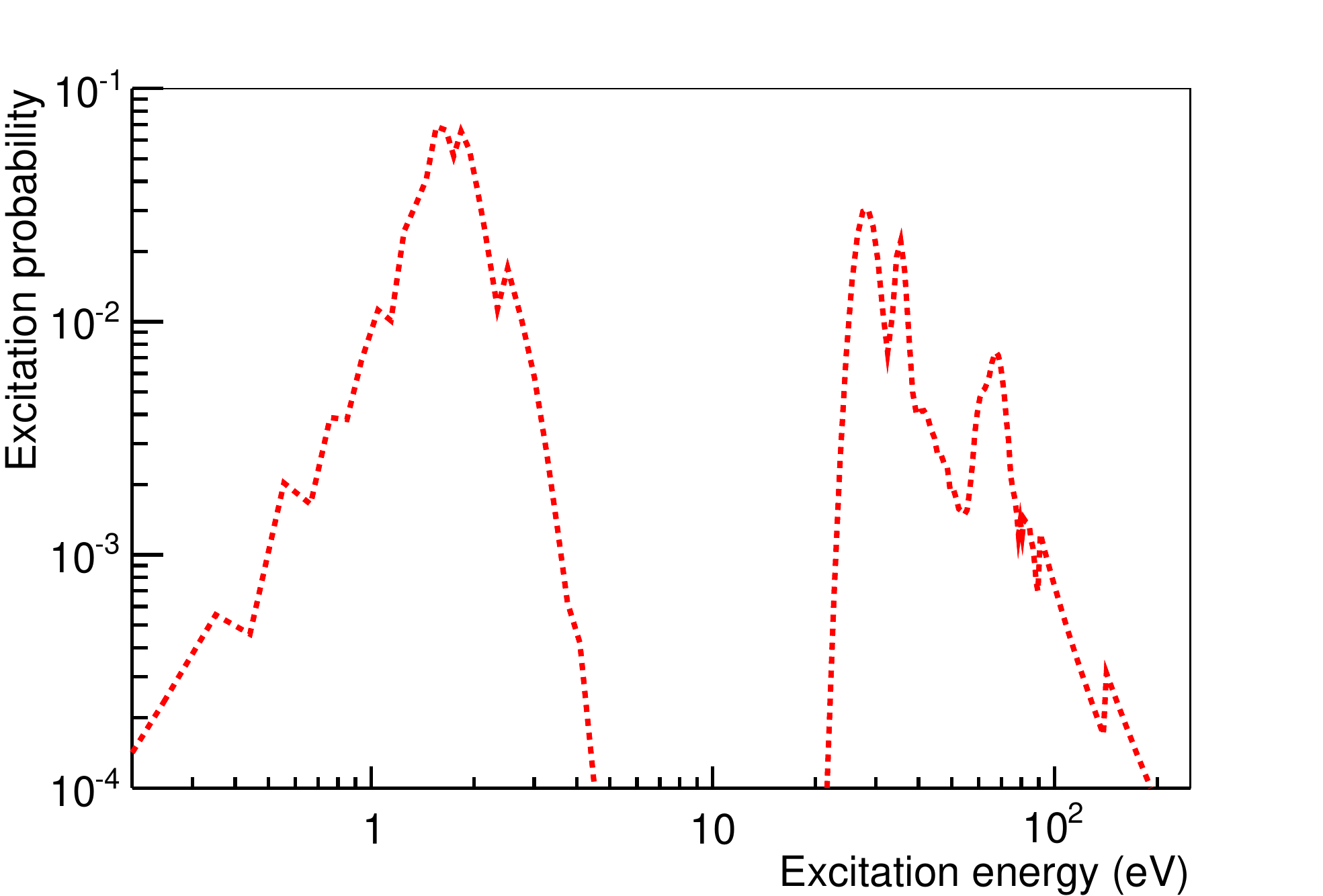}}
  \subfigure[]{\includegraphics[width = 0.49\textwidth]{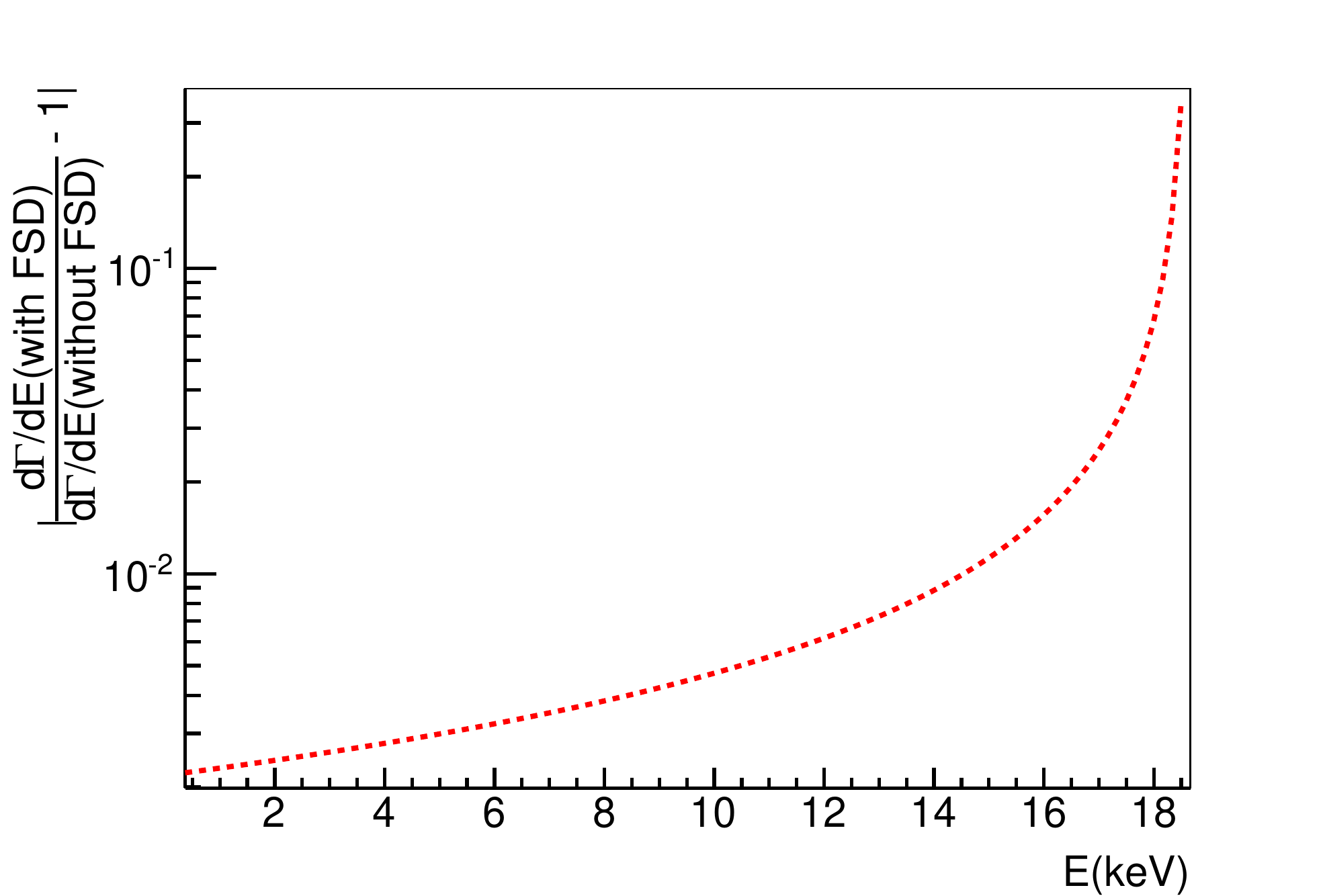}}
  \caption{a) Final State Distribution (FSD) as computed close to the endpoint by~\cite{Saenz2000}. All excitation energies below 5~eV correspond to rotational and vibrational excitations, whereas electronic excitations occur above 20 eV. b) Ratio of tritium $\beta$-decay spectrum with and without the FSD computed close to the endpoint, included as described in equation~\ref{eq:fullbetaspectrum}. As long as the excitation probabilities $P(E,E_{f})$ change smoothly as a function of $\beta$-electron energy $E$, the final state distribution leads to a smooth correction for $E < E_0-200$~eV, since all excitation energies $E_{f}$ are smaller than $200$~eV. }
 \label{fig:FSD}
\end{figure}

\paragraph{Decay into Excited Final States}
After the $\beta$ decay the electron shell of the decaying atom has to rearrange itself into the eigenstates of the daughter atomic ion. Therefore, not only is the atomic ground state populated, but a fraction $P_{f}$ of the decay ends in states with excitation energy $E_{f}$ (see figure~\ref{fig:FSD}). 

In the case of molecular tritium one must consider the decay to electronic excited states of the T$^{3}$He$^+$ daughter molecule, as well as rotational and vibrational excited states. The total decay rate for a single mass eigenstate $m_{\nu}$ is calculated by summing over all of possible final states $f$ of the daughter system with
\begin{small}
\begin{equation}
\label{eq:betaspectrum_fsd}
 \frac{d\Gamma}{dE}(E,E_{0},m_{\nu}) \propto \sum_f P_{f}(E,E_{f})\cdot p \cdot (E + m_e) (E_{0} - E - E_{f})\sqrt{(E_{0} - E - E_{f})^2 - m_{\nu}^2}  \end{equation} 
\end{small}

The effect of the final state distribution (FSD) is significant in the case of molecular tritium. For $\beta$-electron energies close to the endpoint, only 57.4~\% of all decays end in the (rotational-vibrational broadened) ground state of T$^{3}$He$^+$~\cite{Saenz2000,Doss2008}. 
 
So far, the FSD has only been computed for a region close to the endpoint, i.\ e. $P(E,E_{f}) = P(E_0,E_f)$, where it is of interest for determining the light neutrino mass. The calculations are performed using the so-called sudden approximation, which assumes that the outgoing electrons are moving very fast compared to the typical velocity of the orbital electrons and the nuclei in a molecule in the classical picture. Further corrections, beyond the sudden approximation, are small for $E\approx E_0$~\cite{Saenz1996}. For lower electron energies the sudden approximation has to be validated. 

An obvious energy dependence of the FSD arises from the fact that rotational-vibrational excitation gains its energy from the recoil of the decaying atom in the molecule, balancing the recoil of the $\beta$-electron and the anti-neutrino. Thus, the average excitation energy (1.7~eV for $E\approx E_0$) of the rotational-vibrational states will decrease for smaller $\beta$-electron energies.

As a calculation of the FSD as a function of $\beta$-electron energy is not yet available, this correction currently constitutes the largest uncertainty of the tritium $\beta$-decay spectrum. However, a full calculation is expected to be feasible~\cite{SaenzPriv}. In this work we use the FSD as calculated close to the endpoint.

\paragraph{Further Atomic Physics Correction Terms}
The Coulomb interaction between the outgoing electron and the daughter nucleus, described by the relativistic Fermi function F, must be corrected by the screening effect of the orbital electrons left behind from the tritium molecule, which affects the Coulomb field of the daughter T$^{3}$He$^+$ pair of nuclei. This introduces a multiplicative factor S~\cite{Behrens1982}. 

Both terms S and F are almost energy independent close to the endpoint, which is advantageous in the light neutrino mass measurement. However, the effect of both F and S is enhanced at low electron energies, which makes an exact implementation of these terms necessary for the keV-scale sterile neutrino search. In this work we use the screening potential V$_0$ value evaluated for the daughter $^{3}$He$^+$ ion. The exact value of V$_0$ for T$^{3}$He$^+$ molecule will have to be computed precisely for the eventual search for keV-scale sterile neutrinos.

Since the orbital electron of the tritium atom is identical to the emitted electron we must consider the possible quantum mechanical interchange of the $\beta$-particle on emission with the orbital electron. This introduces the multiplicative factor I, which increases steeply as the $\beta$-electron energy decreases. We improved the evaluation of~\cite{Wilkinson1991} which considered only the transition to the 1s ground state (70\% of the decays), by following the prescription of~\cite{Hax85} to include the averaged effect of the decay to the first 10 bounded excited states of $^{3}$He$^+$. Our new parametrization now gathers 97\% of the decays. In~\cite{Wilkinson1991} the magnitude of this correction was overestimated by neglecting the negative interferences induced by the exchange with electrons from $^{3}$He$^+$ excited states. Eventually, the interchange correction will have to be computed for the excited states of T$^{3}$He$^+$ instead of $^{3}$He$^+$. 

\paragraph{Nuclear and Particle Physics Correction Terms}
The T$^{3}$He$^+$ recoil effect must be taken into account since it changes the two-body phase space into a three-body phase space. Weak magnetism and the V-A interference effects alter the form of the spectrum in a similar way. As recommended in~\cite{Bilenkii1960, Mas07} it is convenient to combine the handling of these three effects with a single multiplicative factor R. This term has only a slight positive linear dependence on the electron energy. The effect that the emitted electron does not propagate in the field of a stationary charge but in the field of a charge that is itself recoiling is described by the multiplicative factor Q~\cite{Wilkinson1982}.

The daughter He nucleus is not a point-like particle, and finite-size effects must be accounted for. The radial dependence of the Coulomb field is no longer a $\frac{1}{r^2}$ scaling within the nucleus, leading to a multiplicative correction factor $\mathrm{L}_{0}$~\cite{WilkinsonFS1990}. An evaluation of the electron and the neutrino wave functions appropriately throughout the nuclear volume leads to a further correction C~\cite{WilkinsonFS1990}. Both of these terms have only a slight negative linear dependence with the electron energy.

The effect of radiative corrections are accounted for by the multiplicative term G. In our work, we consider only the order~$\alpha_{\mathrm{em}}$ correction~\cite{Repko1983}. This correction varies rapidly, as $\ln$(E$_0$-E), towards the endpoint. 

The total decay rate for a single mass eigenstate $\text m_{\nu}$ including all multiplicative correction functions $\Psi_i$ = S, I, G, Q, R, C, L$_{0}$ and the FSD is given by
\begin{small}
\begin{equation}
\label{eq:fullbetaspectrum}
 \frac{d\Gamma}{dE}(E,E_{0},m_{\nu}) \propto \sum_f \left[ P_{f}(E,E_{f})\cdot p \cdot (E + m_e) (E_{0} - E - E_{f})\sqrt{(E_{0} - E - E_{f})^2 - m_{\nu}^2} \cdot \prod_{i} \Psi_i(E, E_0) \right].
 \end{equation} 
\end{small}

\begin{table}[h]
\caption{For each multiplicative function $\Psi_i$ ($\Psi_i$ = S, I, G, Q, R, C, L$_{0}$) and the final state distribution (FSD) correction we define $\delta \Psi_i = \frac{(d\Gamma/dE)^{\mathrm{corr}}}{(d\Gamma/dE)^{\mathrm{uncorr}}}-1$. The variation over the whole energy spectrum is defined by $\Delta\mathcal{A}= |\delta  \Psi_i $(1~keV) - $\delta  \Psi_i $(18~keV). $\sigma_{\Psi_i}$ provides a rough estimate of the uncertainty of each physical effects, obtained by varying key parameters or comparing different calculation methods as described in the 6$^{th}$ column. Additionally, the 6$^{th}$ column contains comments about the current status of the computation.}
\centering
{\footnotesize
\begin{tabular*}{\textwidth}{@{\extracolsep{\fill}}crrrrcc}
\hline 
Correction  & E=1 keV & E=9 keV & E=18 keV & $\Delta\mathcal{A}$ & Comment/Error estimation method & Ref. \\
            & [ppm]   & [ppm]   & [ppm]    & [ppm]               &         &      \\
\hline 
\hline 
$\delta$FSD(E)  & 1400  & -635   & -351175 & 352575 & Computed only for the endpoint & \cite{Saenz2000} \\
$\sigma_{FSD}$  & ---   & ---    & ---     & ---    &  &\\
\hline 
\hline 
$\delta$S(E)  & -8850 & -1765  & -995      & 7860   & V$_0$ computed only for $^{3}$He$^+$ ion & \cite{Behrens1982}\\
$\sigma_S$  & 1780 & 360 & 200             &        & $V_0$ varied by $\pm$10\%  &  \\
\hline 
$\delta$I(E) & 2470 & 45  & 10             & 1320   & Excitations computed only for $^{3}$He$^+$ ion & \cite{Hax85} \\
$\sigma_I$  & 1145 & 20 & 5                &        & Diff. between \cite{Wilkinson1991} \& this work & \\
\hline 
$\delta$G(E)  & -6995 & -7270 & -8110      & 1115   & Only first order considered & \cite{Repko1983} \\
$\sigma_G$  & 25 & 260 & 830               &        & Diff. between \cite{Repko1983} \& \cite{Repko1983} approx.&\\
\hline 
$\delta$Q(E)  & -135 & -45 & -30          & 105    & --- & \cite{Wilkinson1982}\\
$\sigma_Q$  & <1 & <1  & <1                 &        & $\lambda_t$ varied by $\pm$1\%  (3$\sigma$)&\\
\hline
$\delta$R(E)  & 875  & 905  & 935          & 60     & --- & \cite{Bilenkii1960} \\
$\sigma_R$  & 5 & 5 & 5              &        & $\lambda_t$ varied by $\pm$1\% (3$\sigma$) &\\
\hline 
$\delta$C(E)  & -105 & -105  & -105        & 1      & --- &  \cite{Wilkinson1991}\\
$\sigma_C$  & 3 & 3  & 3                   &        & R varied by $\pm$5\% &\\
\hline 
$\delta$L$_{0}$(E)  & -20 & -20  & -20     & 1      & --- &  \cite{Wilkinson1991} \\
$\sigma_L$  & 6 & 6  & 6                   &        & R varied by $\pm$5\%  &\\
\hline 
\hline 
\end{tabular*}}
\label{tab:corrections}
\end{table}

\subsection{Sensitivity Studies Based on the Spectral Fit Approach}
To investigate the effect of an uncertainty in the tritium $\beta$-decay spectrum on the sensitivity we parametrize the correction terms, with the parameters $\alpha_j$, and leave them as free nuisance parameters in the fit. These free parameters allow the spectral shape to mimic a keV-scale sterile neutrino signature and thereby alter the sensitivity.

For this study we consider a differential measurement of the tritium $\beta$-decay spectrum (see sections~\ref{subsec:upgrates}, \ref{subsec:statsen}). To evaluate the sensitivity at 90\% CL including the aforementioned physical effects we minimize the following $\chi^2$ function
\begin{equation}
\label{eq:chi2estimator}
\chi^2 = \sum_i \frac{\left( N_{\mathrm{theo}}^{i}-N_{\mathrm{exp}}^{i}(\alpha_j)\right)^2}{\sigma^2_{i}} + \sum_j  \left(\frac{\alpha^{\mathrm{bestfit}}_j-\alpha^{\mathrm{true}}}{\sigma_j}\right)^2,
\end{equation}
where $N_{\mathrm{theo}}^{i}$ are the simulated data in the no-mixing case, $N_{\mathrm{exp}}^{i}$ the expectation for a given mixing scenario involving a keV-scale sterile neutrino, and $\sigma_{i}$ the statistical uncertainty in each energy bin $E_i$. The parameters $\alpha^{\mathrm{bestfit}}_j$ are the fitted nuisance parameters, $\alpha^{\mathrm{true}}_j$ denote the true parameter and ${\sigma_j}$ is the allowed uncertainty of the nuisance parameters.

We include the theoretical correction terms, as described in section~\ref{sec:theoerror}, of the form $1 + \alpha_{\Psi_i} \delta\Psi_i$, leaving their amplitudes $\alpha_{\Psi_i}$ as free parameters. Furthermore, we include the final state distribution, as computed for the endpoint, allowing for an uncertainty in the ground state and excited state probability by parameterizing them as $\mathrm{P}_{\mathrm{ground}} = 57\% - \alpha_{\mathrm{FSD}}$ and $\mathrm{P}_{\mathrm{ex}} = 43\%+\alpha_{\mathrm{FSD}}$.

To incorporate any unknown correction, we include a so-called shape factor in our model of the form:
\begin{equation}
 SF(E,E_{0})=1+\alpha_{\mathrm{pol_1}}\left(\frac{E}{E_{0}}\right)+\alpha_{\mathrm{pol_2}}\left(\frac{E}{E_{0}}\right)^{2}+\alpha_{\mathrm{pol_3}}\left(\frac{E}{E_{0}}\right)^{3}.
\end{equation}
The tritium $\beta$-decay spectrum is multiplied by this factor and the parameters $\alpha_{\mathrm{pol_{i = 1, 2, 3}}}$ are left as free parameters.

\begin{figure}
  \centering
  \begin{minipage}{0.7\textwidth}
    \includegraphics[width = \textwidth]{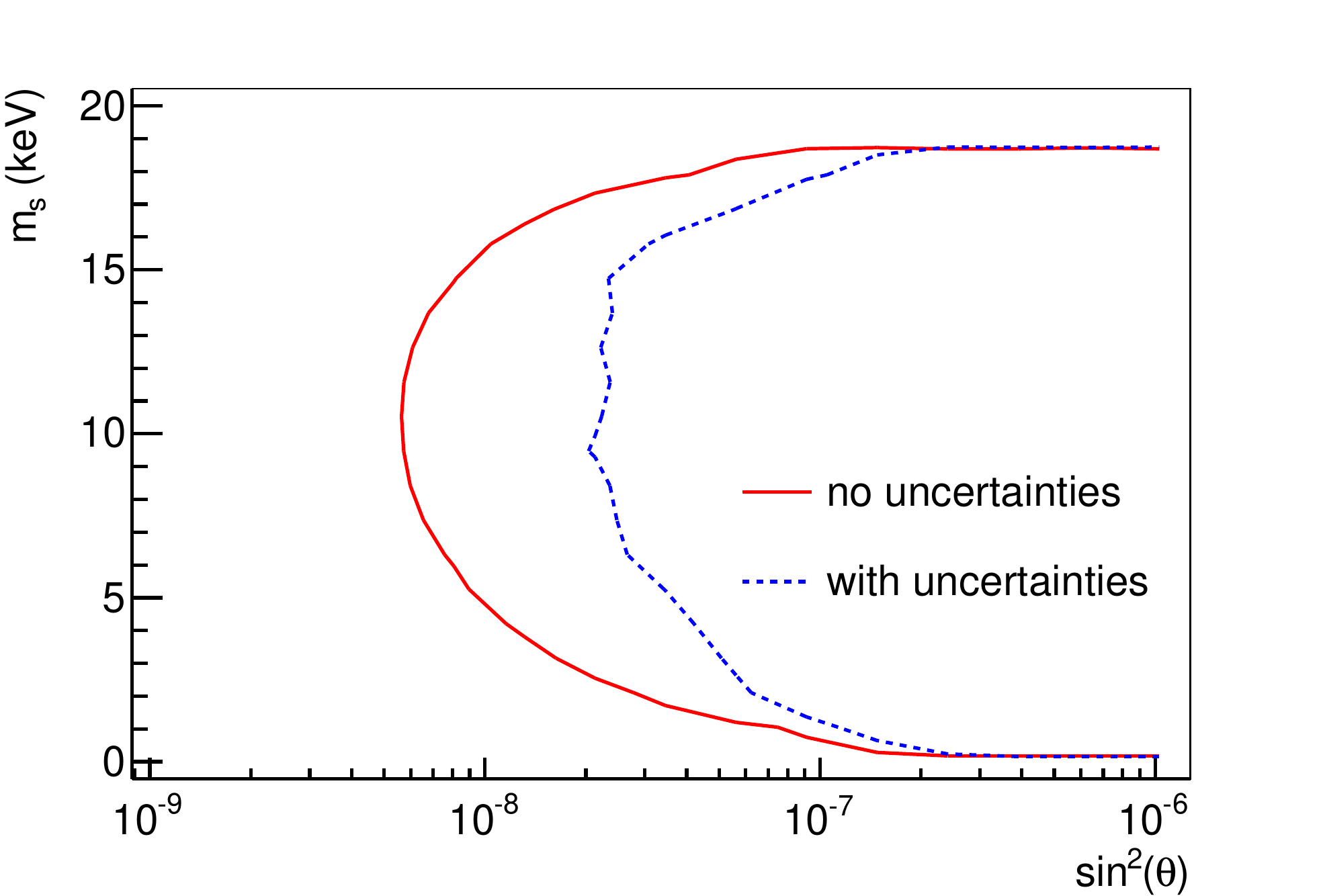}
  \end{minipage}
  \hfill
  \begin{minipage}{0.29\textwidth}
    \caption{90\% exclusion limit of a differential measurement (red solid line) without uncertainties and with uncertainties on the theoretical spectrum. The latter is obtained by a spectral fit approach, allowing for 12 free parameters, describing the theoretical uncertainties to the tritium $\beta$-spectrum (see main text).}
    \label{fig:TheoExclusion}
  \end{minipage}
\end{figure}

We obtain the exclusion curve by performing a 12-free-parameter $\chi^2$-fit at each grid point in the ($m_{\mathrm{s}}, \sin^2\theta$) plane. As a result, we find that the maximal sensitivity is approximately reduced by a factor of about 5 compared to the case of no theoretical uncertainties, see figure~\ref{fig:TheoExclusion}. In other words, the physical effects considered here can not mimic a kink for mixing amplitudes of $\sin^2\theta \gtrapprox 10^{-7}$. 

It should be emphasized that necessary parameterizations of the tritium $\beta$-decay spectrum are required in order to perform a fit and realize this analysis approach. Neglecting certain corrections in the model of the $\beta$-decay spectrum can lead to poor fit results when fitting to the MC data including those corrections.

\section{Impact of Experimental Uncertainties}
\label{sec:sysex}
In this section we consider the impact of detector energy resolution and detection efficiency. The latter is used to study the effect of correlated uncertainties using the covariance matrix approach. In this study we assume a differential measurement with a silicon based detector.  

\subsection{Detector Energy Resolution}
\begin{figure}
  \centering
  \subfigure[]{\includegraphics[width = 0.45\textwidth]{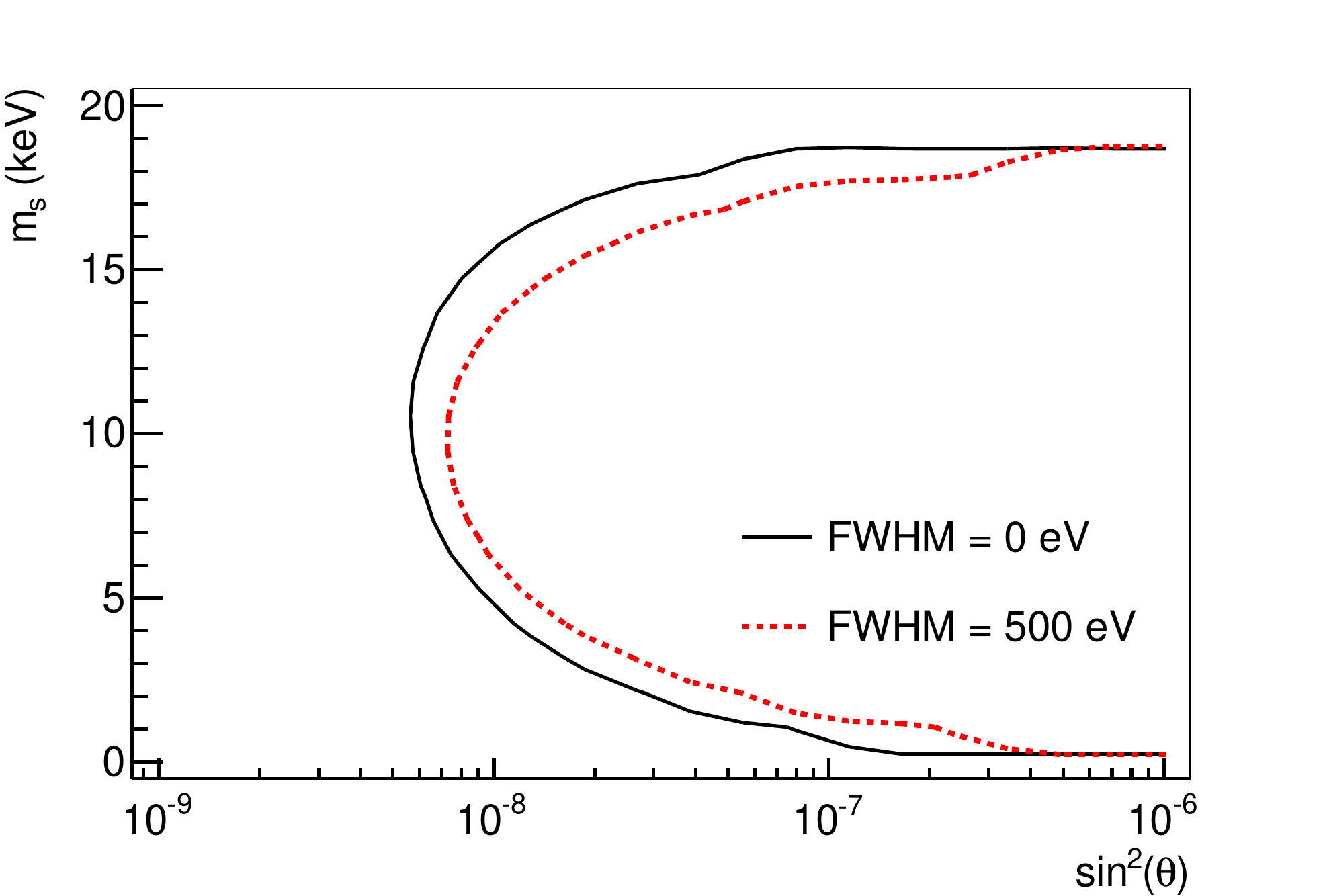}}
  \subfigure[]{\includegraphics[width = 0.45\textwidth]{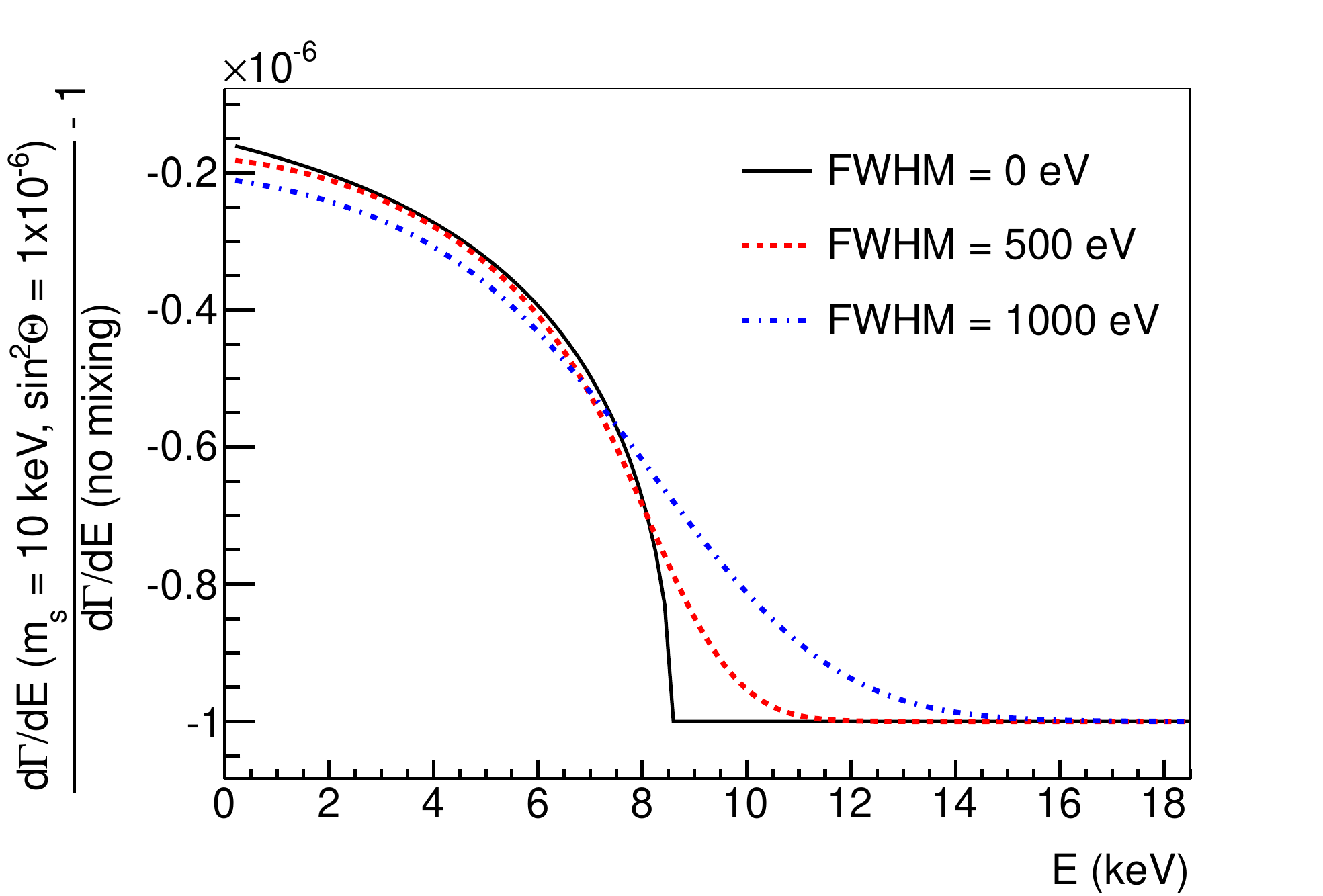}}
  \caption{a) 90\% exclusion limit of a differential measurement with a detector energy resolution of 500~eV FWHM. b) Ratio of spectra with $m_{\mathrm{s}}=10$~keV and $\sin^2\theta=10^{-5}$. $\mathrm{FWHM}=0$~eV (black solid line), $\mathrm{FWHM} = 500$~eV (red dashed line) and $\mathrm{FWHM} = 1000$~eV (blue dashed pointed line).}
 \label{fig:Eres}
\end{figure}

The detector resolution is implemented as a convolution of the spectrum with a Gaussian distribution. A finite energy resolution smooths out the kink signature, as shown in figure~\ref{fig:Eres}. In this general consideration we neglect non-Gaussian tails in the energy resolution. 

The Fano factor for silicon is about 0.12. This implies that, for an electron of 20~keV energy, the charge production statistics contribute only about 220~eV to the energy FWHM. We consider a reasonable electronic noise contributions of FWHM$_{\mathrm{noise}}$ = 450~eV. With this assumption, the total FWHM 
\begin{equation}
 \mathrm{FWHM} = \sqrt{\mathrm{FWHM}_{\mathrm{noise}}^2(\mathrm{eV}^2) + 2.42~\mathrm{eV} \cdot E(\mathrm{eV})},
\end{equation}
changes by at most 10\% over the whole energy range. We have therefore conservatively chosen to use the $\mathrm{FWHM}(20~\mathrm{keV})=500~\mathrm{eV}$ for all energies. 

In the spectral fit approach, the sensitivity is not significantly reduced with a detector resolution of up to 500~eV FWHM. One can understand this feature from the fact that the signature of a keV-scale sterile neutrino is not a purely local effect at $E = E_0 - m_{\mathrm{s}}$. Its impact on the spectrum extends from the kink down to very low energies. The spectral fit approach makes use of the entire information offered by the spectral shape. In contrast, an algorithm searching for the local effect of the kink can require a less precise knowledge of the spectrum but demands a high energy resolution~\cite{wavelet}.

\subsection{Detection Efficiency}
In this section we investigate the sensitivity with respect to experimental uncertainties that are correlated over various energy ranges. A number of experimental effects, such as energy non-linearity due to detector dead-layer and backscattering effects, digitizer differential non-linearity, energy-dependent detection efficiency, and source-related effects such as energy loss due to inelastic scattering, will lead to uncertainties which are correlated over varying energy ranges. 

\begin{figure}
  \centering
  \begin{minipage}{0.49\textwidth}
    \includegraphics[width = \textwidth]{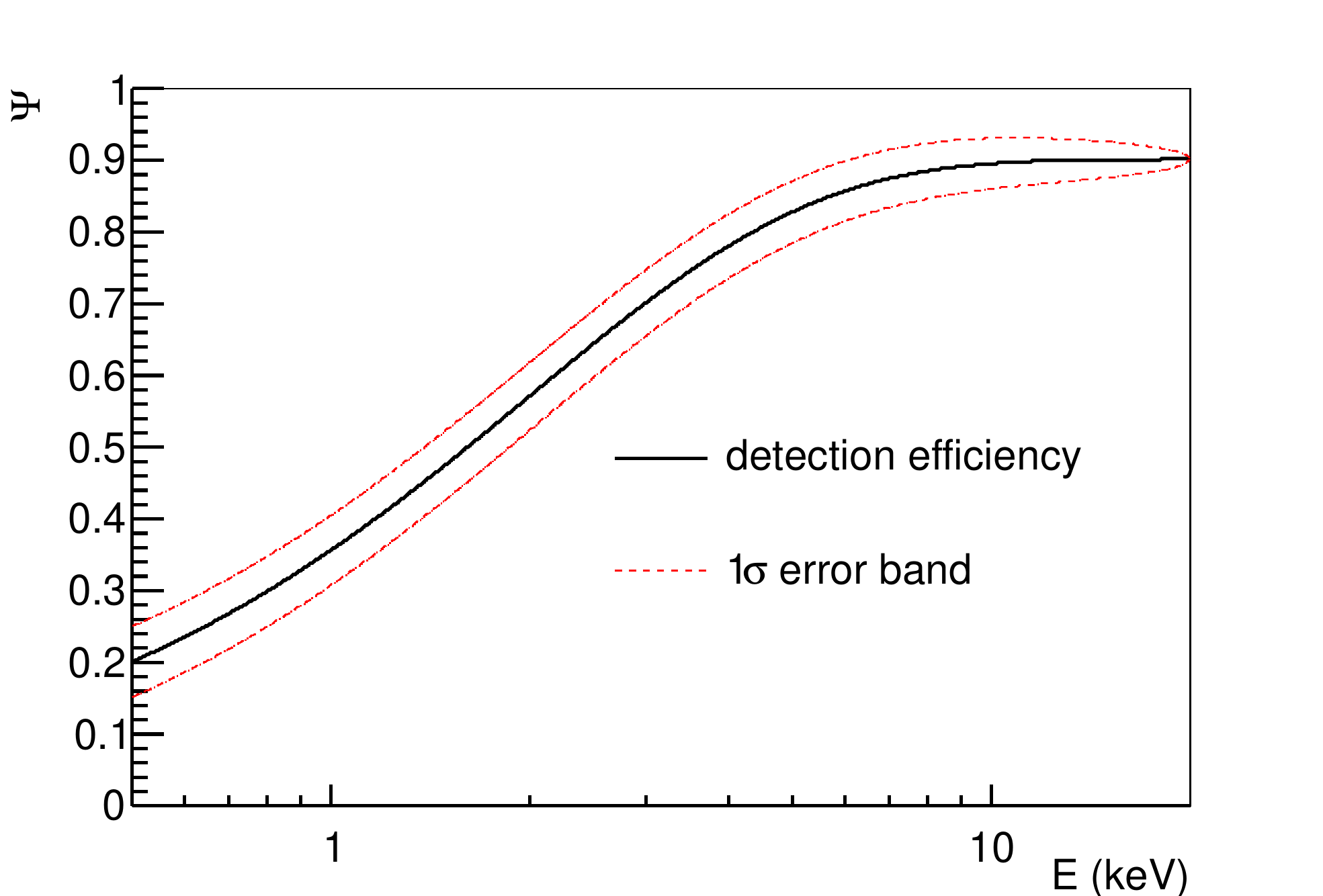}
  \end{minipage}
  \hfill
  \begin{minipage}{0.49\textwidth}
    \caption{Detection efficiency as a function of energy (according to equation~\ref{eq:deteff}) with $\epsilon _{\mathrm{max}}=90\%$ and $x=10$. The red dashed line indicates the energy-dependent error (according to equation~\ref{eq:detefferror}) for $\rho$=5$\%$.}
    \label{fig:Deff}
  \end{minipage}
\end{figure}

In the following we use the detection efficiency as a case study. Regardless of the technology utilized in the future, the detection efficiency will likely be measured through extensive calibration campaigns, using well-known electron sources at specific electron energy intervals. Since the detection efficiency is expected to be a continuous, smooth function of energy, the resulting measured efficiency will have an uncertainty that is correlated over a certain energy range. We assume a generic energy-dependent detection efficiency of the form 
\begin{equation}
\label{eq:deteff}
 \Upsilon(E) = \epsilon _{\mathrm{max}} \cdot ( 1 - e^{-x\cdot\frac{E}{E_0}}),
\end{equation}
where $\epsilon _{\mathrm{max}}$ denotes the maximum efficiency reached at a certain energy. In this work we assume a maximum detection efficiency of 90\% and we conservatively estimate a reduction of the efficiency to 20\% at 500~eV. Furthermore, we assume an energy-dependent uncertainty in the detection efficiency of the form
\begin{equation}
\label{eq:detefferror}
 \delta \Upsilon(E) = \rho \cdot \left(1-\frac{E}{E_0}\right)^{\frac{1}{2}},
\end{equation}
where $\rho$ determines the overall size of the uncertainty. Figure~\ref{fig:Deff} shows the detection efficiency with its 1$\sigma$ error band assuming Gaussian tails. Making use of a post-acceleration electrode (PAE), installed at the exit of the main spectrometer, that boosts the kinetic energy of all $\beta$-decay electron by $qV_{\mathrm{PAE}}$ may improve the detection efficiency significantly. 

\begin{figure}
  \centering
  \subfigure[]{\includegraphics[width = 0.49\textwidth]{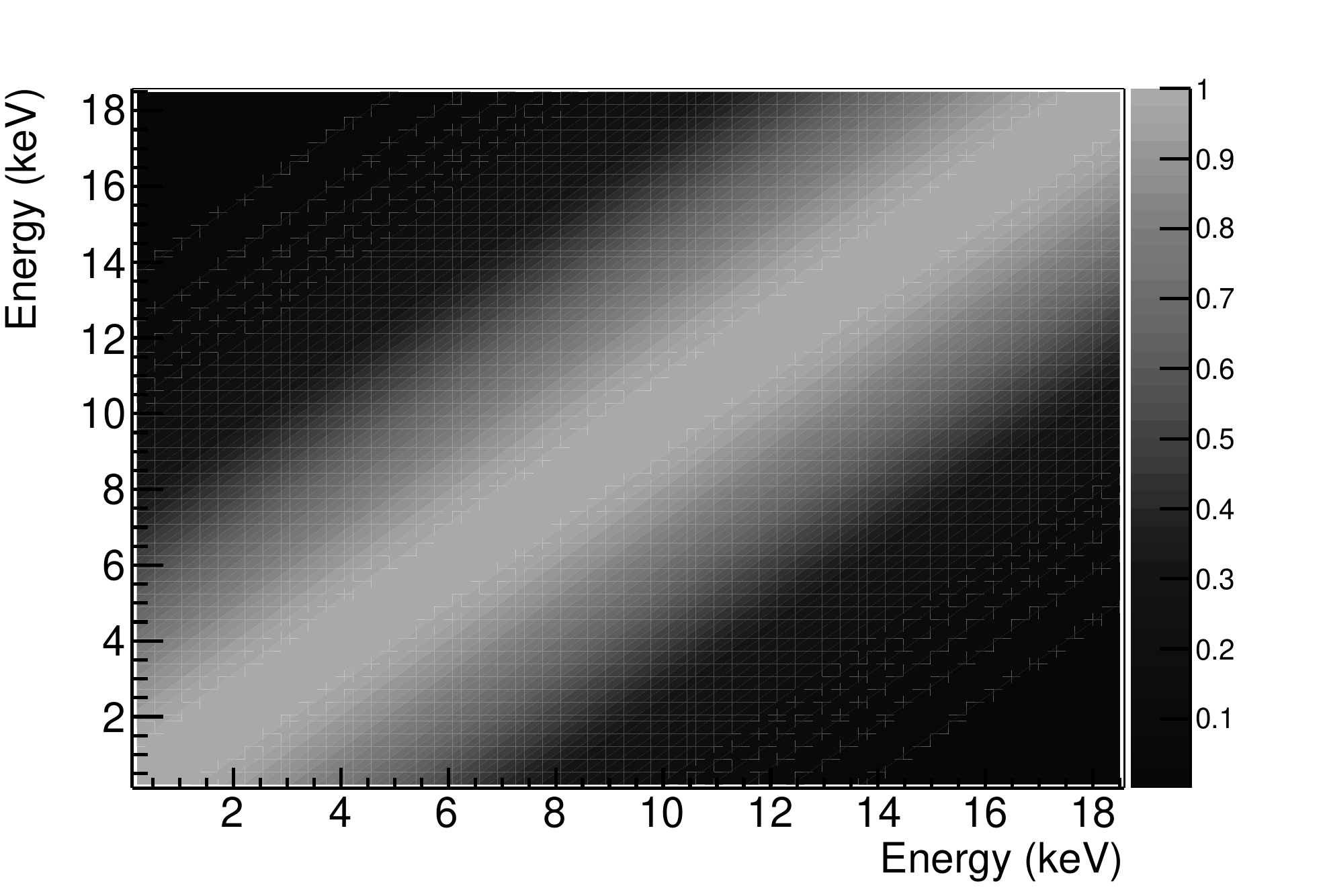}}
  \subfigure[]{\includegraphics[width = 0.49\textwidth]{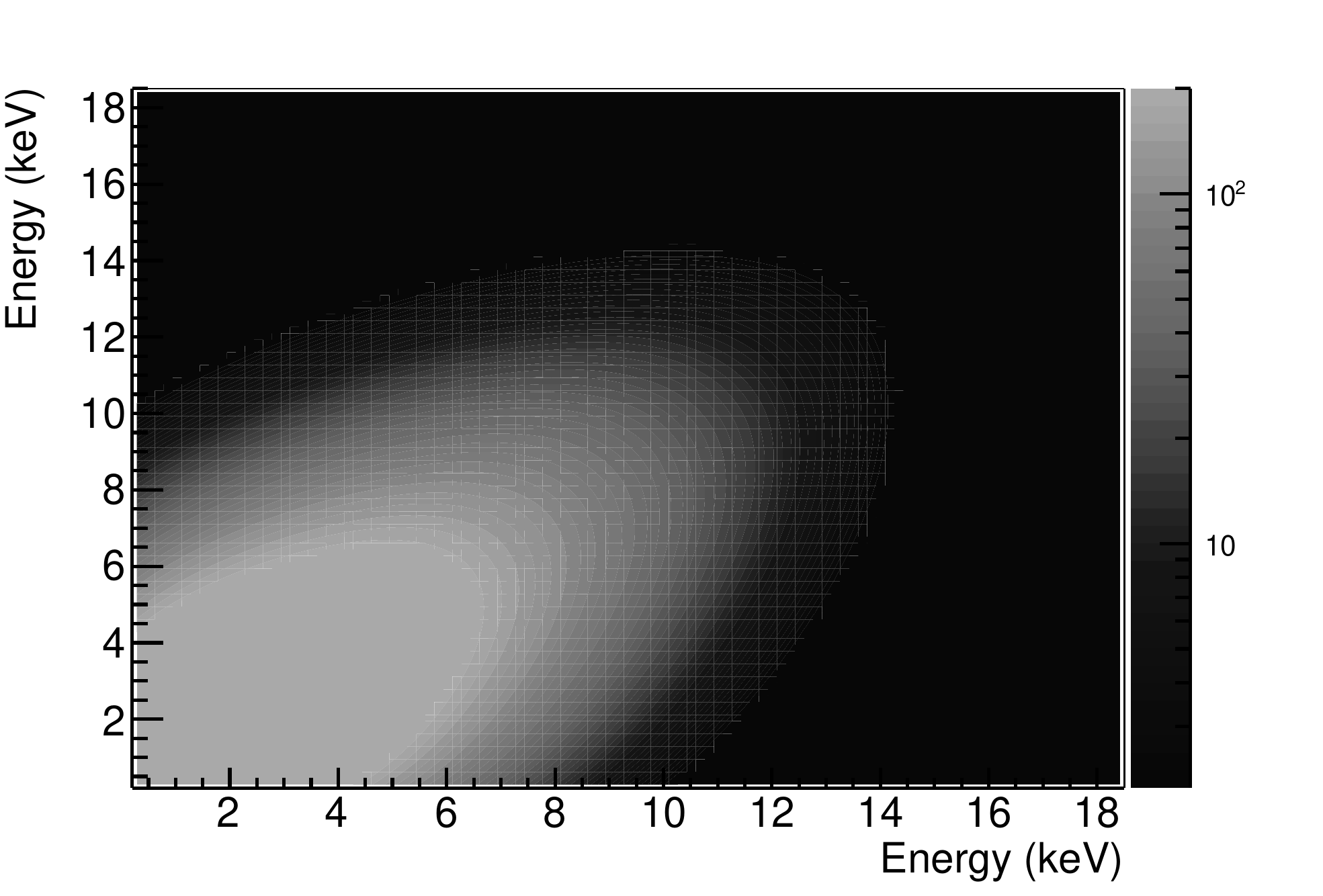}}
  \caption{a) Weighting factors $\omega_{ij}$ expressing the error correlation length for $\Delta E = 5$~keV. b) Covariance matrix as defined in equation~\ref{eq:covmat} for $\rho=0.5\%$.}
 \label{fig:matrix}
\end{figure}

In order to incorporate the effect of correlation we introduce a covariance matrix $V_{ij}$. The covariance matrix contains the fully uncorrelated statistical uncertainty $V_{ij}^{\mathrm{stat}}$ and a correlated error represented by the matrix $V_{ij}^{\mathrm{sys}}$,
\begin{equation}
\begin{array}{lcll}
         V_{ij}^{\mathrm{stat}} & = & N_{\mathrm{exp}}^{i} & \mbox{ for i=j} , \\
         V_{ij}^{\mathrm{sys}}  &=  & \omega_{ij} \cdot N_{\mathrm{exp}}^{i}\cdot N_{\mathrm{exp}}^{j} \cdot \delta \Upsilon^{i} \cdot \delta \Upsilon^{j}& \mbox{ all i and j}.\\
\end{array}
\label{eq:covmat}
\end{equation}
To investigate the impact of a different degree of correlation we introduce a parameter $\omega$ which weights the off-diagonal terms in the covariance matrix according to a Gaussian distribution. Its width $\Delta E$ defines the correlation length, i.e. over which energy range the data points are correlated. Figure~\ref{fig:matrix} shows the weighting parameter $\omega$ for a width of $\Delta E = 5$~keV and the corresponding covariance matrix.

We obtain the exclusion curve of a 3-years differential measurement by computing the $\chi^2$ as follows:
\begin{equation}
\label{eq:chi2estimator2}
\chi^2 = \sum_{ij}  \Upsilon_i \left( N_{\mathrm{theo}}^{i}-N_{\mathrm{exp}}^{i}(\alpha_k) \right)  V_{ij}^{-1} \left( N_{\mathrm{theo}}^{j}-N_{\mathrm{exp}}^{j}(\alpha_k)\right)\Upsilon_j + \sum_k  \left(\frac{\alpha^{\mathrm{bestfit}}_k-\alpha^{\mathrm{true}}}{\sigma_k}\right)^2. 
\end{equation}
$\chi^2$ is computed at each grid point in the ($m_{\mathrm{s}}, \sin^2\theta$) plane, while fixing all theoretical parameters to their expected values, keeping only the overall signal count rate as an unconstrained free parameter. 

\begin{figure}[h!]
  \centering
  \begin{minipage}{0.7\textwidth}
    \includegraphics[width = \textwidth]{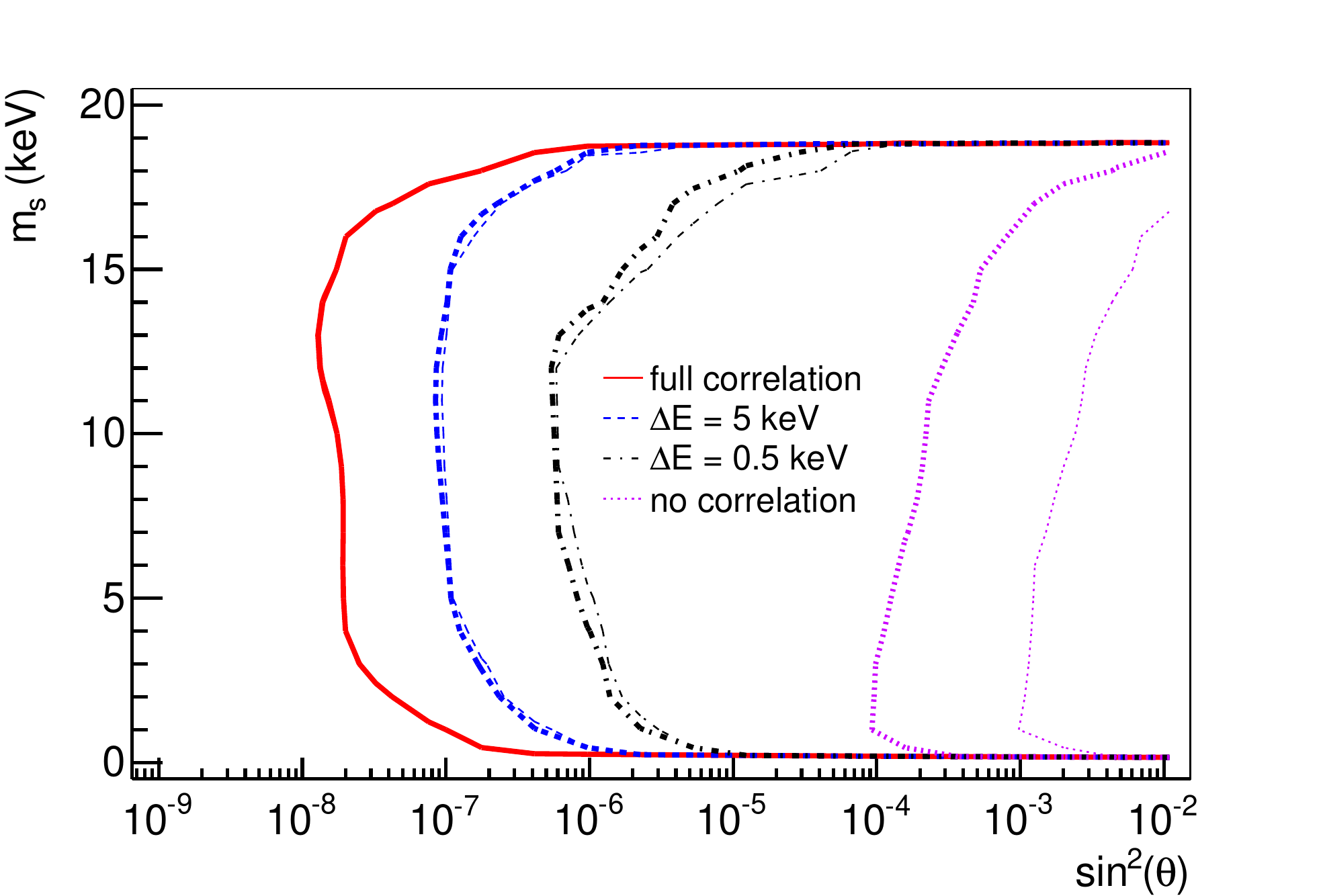}
  \end{minipage}
  \hfill
  \begin{minipage}{0.29\textwidth}
	\caption{Exclusion curves for different correlation lengths $\Delta E$ and different sizes of systematic uncertainties $\rho$. The thin lines correspond to a value of  $\rho = 0.5\%$, the bold lines to $\rho = 0.05\%$.}
	\label{fig:Deff_exclusion}
  \end{minipage}
\end{figure}

Figure~\ref{fig:Deff_exclusion} shows the sensitivity for different efficiency uncertainties $\rho$ and different correlation lengths $\Delta E$. The sensitivity depends strongly on the correlation length. In the case of a fully uncorrelated error of the order of $\rho=10^{-3}$ the sensitivity is reduced to $\sin^2\theta \approx 1\cdot10^{-3}$. Considering a more realistic case including a correlation over an energy range of $\Delta E = 0.5$~keV, the result is improved and interestingly, the dependence on $\rho$ is strongly reduced. For a correlation length of $\Delta E = 5$~keV, a sensitivity of $\sin^2\theta \approx 1\cdot10^{-7}$ can be reached for $\rho$ as large as a few percent. 

We conclude that any uncorrelated systematic effects will strongly affect the final sensitivity. Precise calibrations, simulations, and understanding of correlations will be essential when performing a differential measurement of tritium $\beta$-decay spectrum to hunt for keV-scale sterile neutrinos. As this approach is general in its nature, this conclusion holds true for a broad class of systematic effects. Specific studies of the systematic uncertainties associated with the details of the future technical realization of this measurement, utilizing more sophisticated statistical methods, will have to be performed to deduce the final sensitivity.

\section{Discussion and Outlook}
In this work we show that a three-year measurement of the tritium $\beta$-decay spectrum with a source strength similar to that of the KATRIN experiment will reach a statistical sensitivity to keV-scale sterile neutrinos with mixing amplitudes down to $\sin^2\theta = 10^{-8}$ (90\% C.L.). Hence, from a purely statistical point of view this would allow the probing of the parameter space of significant cosmological interests in a model-independent laboratory measurement.

To assess a realistic sensitivity, we investigated the effect of theoretical uncertainties on the tritium $\beta$-decay spectrum with a spectral fit approach. In doing so, we implemented a state-of-the-art description of the tritium $\beta$-decay spectrum, allowing for uncertainties in the parameters describing the corrections to the tritium $\beta$-decay spectrum. These investigations show that smooth corrections do not mimic a kink with $\sin^2\theta > 10^{-7}$. However, a complete parametrization of all corrections to the $\beta$-spectrum will be necessary to perform a spectral fit and therefore a more complete calculation of several theoretical corrections, such as the final state distribution, will be mandatory.

We included ubiquitous experimental effects such as finite detector resolution and energy-dependent detection efficiency and find that an energy resolution of up to FWHM = 500~eV does not reduce the sensitivity significantly. The detection efficiency serves as an example of a partly correlated systematic uncertainty. This case study shows that any small fraction of uncorrelated uncertainties will strongly affect the final sensitivity. 

The signature of a keV-scale sterile neutrino on the shape of the $\beta$-decay spectrum is not only a local kink-like effect but extends over a wide energy range. This characteristic spectral modification will be important to distinguish a possible signal from systematic effects. Furthermore, combining the differential, integral and TOF measurement schemes will be valuable in probing a large set of experimental systematic uncertainties, and can help to uncover false positive signals. Eventually, independent laboratory experiments based on different isotopes or astrophysical X-ray observations should confirm a positive signal.

Finally, we would like to emphasize that an extensive study of all individual experimental systematic effects needs to be performed to assess the final sensitivity. These studies will require a detailed description of the technical realization of the tritium source, transport and detection system of a future direct keV-scale sterile neutrino search experiment.

\section*{Acknowledgements}
S. Mertens gratefully acknowledges support of a Feodor Lynen fellowship by the Alexander von Humboldt Foundation and support by the Helmholtz Association. This work was supported by the U.S. Department of Energy, Office of Science, Office of Nuclear Physics, under Contract No. DE-AC02-05CH11231. S. Groh and A. Huber would like to thank KIT and KHYS for financial support and LBNL for hospitality during their research stay at Berkeley. Th.\ Lasserre thanks the European Research Council for support under the Starting Grant StG-307184. N. Steinbrinck acknowledges funding by BMBF grant 05A11PM2. Special thanks to D.\ Radford and S.\ Neubauer for fruitful discussions.


\appendix
\section{Theoretical Correction Terms}
\label{appendix}
In this section we show the mathematical description of all correction terms used in this work. The electron energy is expressed in units of the electron mass $W=(E+m_{e})/m_{e}$. \\

\paragraph*{Relativistic Fermi Correction} 
\begin{equation}
 F(Z,W)=4(2pR/m_{e})^{-2(1-\gamma)}\cdot \left|\Gamma(\gamma+iy)\right|^{2}/\left(\Gamma(2\gamma+1\right))^{2}\cdot  e^{\pi\gamma},
\label{eq:corrF}
\end{equation}
where Z=2 for $^{3}$He, $\alpha_{\mathrm{em}}$ is the fine structure constant, $y=\alpha_{\mathrm{em}} ZW/p$,$\gamma=(1-\alpha_{\mathrm{em}}^{2}Z^{2})^{0.5}$, and $R=2.8840\cdot 10^{-3}$
is the nuclear radius of the $^{3}$He nucleus, in units of m$_{e}$, given by the Elton formula~\cite{Elton1958}. \\

\paragraph*{Screening of the Nuclear Potential by the Orbital Electron} 
\begin{equation}
 S(Z,W)=\overline{W}/W\left(\overline{p}/p\right)^{2\gamma-1}exp\left(\pi(\overline{y}-y)\right)\frac{\left|\Gamma(\gamma+i\overline{y})\right|^{2}}{\left|\Gamma(\gamma+iy)\right|^{2}},
\label{eq:corrZ}
\end{equation}
where $\overline{y}=\alpha_{\mathrm{em}} Z\overline{W}/p$, $\overline{W}=W-V_{0}$, and $V_{0}\sim1.45\alpha_{\mathrm{em}}^{2}$ denote the screening potential~\cite{Behrens1982}. \\

\paragraph*{Quantum Mechanical Effect of an Orbital Electron Exchange}
\cite{Hax85}
\begin{equation}
 \delta I(Z,W)=2.462\cdot \alpha(\eta)^{2}+0.905\cdot \alpha(\eta),
\label{eq:corrE}
\end{equation}
where $\alpha(\eta)=\eta^{4}exp\left[2\eta\cdot artan(-2/\eta)\right]/(1+0.25\eta^{2})^{2}$,
with $\eta=-2\alpha_{\mathrm{em}}/p$.

\paragraph*{Recoil Effects, Weak Magnetism and V-A Interference} 
\cite{Bilenkii1960}
\begin{equation}
 \delta R(W,W_{0},M)=(a\cdot W-b/W)/c,
\label{eq:corrR}
\end{equation}
where $a=2(5\lambda_{t}^{2}+2\lambda_{t}\cdot \mu+1)/M$, $b=2\lambda_{t}(\lambda_{t}+\mu)/M$,
$c=1+3\lambda_{t}^{2}-b\cdot W_{0}$. M denotes the mass of the $^3$He in units of $m_{e}$. The ratio of 
the axial-vector to the vector coupling constant for weak interaction in triton beta decay is
 $\lambda_{t}=1.265 \pm 0.0035$~\cite{Akulov2002}. The difference between the total magnetic moments of 
the triton and helion is $\mu$=5.106588, with a negligible uncertainty~\cite{nistwww}.

\paragraph*{Recoiling Coulomb Field}
\cite{Wilkinson1982}
\begin{equation}
 \delta Q(Z,W,W_{0})=\frac{\pi\cdot \alpha_{\mathrm{em}}\cdot Z}{M\cdot (p/m_e)}\left(1+\frac{1-\lambda_{t}^{2}}{1+3\lambda_{t}^{2}}\frac{W_{0}-W}{3W}\right). 
\label{eq:corrQ}
\end{equation}

\paragraph*{Finite Extension of the Nucleus} 
\cite{WilkinsonFS1990}
\begin{equation}
 \delta L_{0}(Z,W)=\frac{13}{60}\cdot (\alpha_{\mathrm{em}} Z)^{2}-W\cdot R\cdot \alpha_{\mathrm{em}} \cdot Z\cdot \frac{41-26\gamma}{15(2\gamma-1)}-\alpha_{\mathrm{em}}\cdot Z\cdot R\cdot \gamma\cdot \frac{17-2\gamma}{30\cdot W(2\gamma-1)},  
\label{eq:corrL0}
\end{equation}
and 
\begin{equation}
 \delta C(Z,W)=C_{0}+C_{1}\cdot  W+C_{2}\cdot  W^{2}, 
\label{eq:corrC}
\end{equation}
with 
\begin{eqnarray}
 C_{0}&=&-\frac{233(\alpha_{\mathrm{em}}\cdot Z)^{2}}{630}-\frac{(W_{0}\cdot R)^{2}}{5}+\frac{2(W_{0}\cdot R\cdot \alpha_{\mathrm{em}}\cdot Z)}{35},\\
 C_{1}&=&-\frac{21(R\cdot \alpha_{\mathrm{em}}\cdot Z)}{35}+\frac{4(W_{0}\cdot R^{2})}{9},\\
 C_{2}&=&-\frac{4R^{2}}{9}.
\label{eq:corrCbis}
\end{eqnarray}

\paragraph*{Radiative Corrections}  
\begin{equation}
G(W,W_{0})=(W_{0}-W)^{(2\alpha_{\mathrm{em}}/\pi)t(\beta)} \left(1+\delta G(W,W_{0})\right),   
\end{equation}
with
\begin{small}
\begin{eqnarray}
\delta G(W,W_{0})&=&\frac{2\alpha_{\mathrm{em}}}{\pi}\left(t(\beta)\left(\ln(2)-\frac{3}{2}+\frac{W_{0}-W}{W}\right) \right. \\
&& \left. +\frac{1}{4}(t(\beta)+1)\left(2(1+\beta^{2})-2\ln(\frac{2}{1-\beta})+\frac{(W_{0}-W)^{2}}{6W^{2}}\right)\right.\\
&& \left. +\frac{1}{2\beta}\left(L(\beta)-L(-\beta)+L(\frac{2\beta}{1+\beta})+\frac{1}{2}L(\frac{1-\beta}{2})-\frac{1}{2}L(\frac{1+\beta}{2})\right)\right),
\label{eq:corrG}
\end{eqnarray}
\end{small}
$t(\beta)=\frac{1}{2\beta}\ln(\frac{1+\beta}{1-\beta})-1$ and
$L(\beta)=\int_{0}^{\beta}\ln(1-t)/t\ dt$, where $\beta=(W^2-1)^\frac{1}{2}/W$. More details can be found in~\cite{Repko1983}.

\end{document}